\documentclass[useAMS,usenatbib]{mn2e}

\usepackage{graphicx}        

\usepackage{multirow}

\newcommand{\apj}{\mbox{ApJ}}

\newcommand{\apjl}{\mbox{ApJL}}

\newcommand{\aj}{\mbox{AJ}}
\newcommand{\mnras}{\mbox{MNRAS}}

\newcommand{\nat}{\mbox{Nature}}

\def\apgt{\ {\raise-.5ex\hbox{$\buildrel>\over\sim$}}\ }
\def\aplt{\ {\raise-.5ex\hbox{$\buildrel<\over\sim$}}\ }
\def\lt{\ {\raise-.5ex\hbox{$\buildrel>$}}\ }
\def\gt{\ {\raise-.5ex\hbox{$\buildrel<$}}\ }

\def\Msun{\ensuremath{{\rm M}_{\odot}}}

\title[The Effect of Many Minor Mergers]{The Effect of Many Minor Mergers on the Size Growth of Compact Quiescent Galaxies}

\author[J. B\'edorf and S. Portegies Zwart]
{J. B\'edorf$^{1}$\thanks{E-mail:
bedorf@strw.leidenuniv.nl (JB)} and S. Portegies Zwart$^{1}$\\
$^{1}$Leiden Observatory, Leiden University, P.O. Box 9513, 2300 RA Leiden, 
The Netherlands}
\begin{document}


\maketitle

\label{firstpage}

\begin{abstract}
Massive galaxies with a half-mass radius $\aplt 1$\,kpc are observed
in the early universe ($z\apgt2$), but not in the local universe.  In
the local universe similar-mass (within a factor of two) galaxies tend
to be a factor of 4 to 5 larger. Dry minor mergers are known to drive
the evolution of the size of a galaxy without much increasing the
mass, but it is unclear if the growth in size is sufficient to explain
the observations.  We test the hypothesis that galaxies grow through
dry minor mergers by simulating merging galaxies with mass ratios of
$q=$1:1 (equal mass) to $q=$1:160. In our $N$-body simulations the
total mass of the parent galaxy doubles.  We confirm that major
mergers do not cause a sufficient growth in size. The observation can
be explained with mergers with a mass ratio of $q=$1:5--1:10. Smaller
mass ratios cause a more dramatic growth in size, up to a factor of
$\sim 17$ for mergers with a mass ratio of 1:80.  For relatively
massive minor mergers $q\apgt $1:20 the mass of the incoming child
galaxies tend to settle in the halo of the parent galaxy.  This is
caused by the tidal stripping of the child galaxies by the time they
enter the central portion of the parent.  When the accretion of minor
galaxies becomes more continuous, when $q\aplt $1:40, the foreign mass
tends to concentrate more in the central region of the parent
galaxy. We speculate that this is caused by dynamic interactions between the child
galaxies inside the merger remnant and the longer merging times when
the difference in mass is larger. These interactions cause dynamical
heating which results in accretion of mass inside the galaxy core and
a reduction of the parent's circular velocity and density.

\end{abstract}


\section{Introduction}

Hierarchical structure formation drives the growth of mass and size of
galaxies with redshift \citep{1980lssu.book.....P,
  1993MNRAS.262..627L}. In this picture the mass and size of galaxies
grow at a similar rate. Recently a population of small but relatively
massive elliptical galaxies has been observed at $z\apgt 2$
\citep{2005ApJ...626..680D,2006ApJ...650...18T,2007ApJ...671..285T,
2008ApJ...677L...5V, 2008ApJ...688..770F, 2011ApJ...736L...9V,
  2012ApJ...749..121S}. However at low redshift these galaxies are
much more rare \citep{2009ApJ...692L.118T, 2009Natur.460..717V, 
2008ApJ...688...48V, 2010ApJ...720..723T, 2010ApJ...721L..19V, 2012ApJ...751...45T},
whereas there is a rich population of elliptical galaxies with similar
mass but which are considerably larger in size
\citep{2011ApJ...738L..22M}.  This suggests that small and massive
galaxies grow in size without acquiring much mass, which cannot be
explained from major mergers as observed in hierarchical structure
formation simulations (e.g. \cite{2006ApJ...636L..81N, 
2012ApJ...744...63O} and references therein).

It has been suggested that dry minor mergers can cause a considerable
growth in size without much increasing the mass of the galaxy, whereas
major mergers tend to increase the mass without much increasing the
size \citep{1980ApJ...235..421M, 2006ApJ...636L..81N, 
2010ApJ...725.2312O,2012ApJ...744...63O, 2011MNRAS.415.3903T}.

Several studies on the topic seem to contradict each other, some argue
that minor mergers can drive the observed growth
e.g. \citep{2009ApJ...697.1290B, 2009ApJ...699L.178N,
  2009ApJ...691.1424H, 2010ApJ...724..915H}. Others claim that the observed 
size growth in simulations is not sufficient if one takes into
account cosmological scaling relations 
\citep{2009ApJ...706L..86N,2012MNRAS.422.1714N,2012MNRAS.422L..62C}.
Encouraged by this discrepancy in the literature we decided to perform
a series of simulations in which we model the encounters between
compact massive galaxies and smaller, lower-mass counterparts.  In our
parameter search we included mass ratios from 1:1 all the way to
1:160. Even though such low-mass encounters are unlikely to be common
\citep{2012ApJ...744...63O}, they approach a continuous in-fall of
material on the parent galaxy.

With the merger simulations we study the growth in mass, size and the
effect on the shape of the merger product. Our simulations are
performed using the {\tt Bonsai} tree code \citep{2012JCoPh.231.2825B} 
with up to 17.2 million equal-mass particles.  This includes the 
dark-matter as well as the baryonic matter.
In our simulations we recognize two regimes of galaxy growth.  For
$q\apgt $1:20 we confirm the inside-out growth, as discussed by
\citep{2012arXiv1206.5004H},
whereas for $q\apgt $1:40 galaxies tend to form
outside-in, meaning that the mass is accreted in the core of the
primary galaxy rather than accreted on the outside. 
This change in behavior is caused by the self-interactions of
minor galaxies for mass ratios $ \aplt 1:40$. 

Recently \citep{2012arXiv1206.1597H} studied the size increase from
major mergers down to 1:10 minor mergers. Their results are consistent
with our findings for comparable simulation parameters.  The major
difference between \citep{2012arXiv1206.1597H} and the results
presented here are the details regarding the orbital parameters of the
merging galaxies; they drop their minor galaxies in one-by-one,
whereas we initialize all galaxies in a spherical distribution at the
start of the simulation. As a consequence in some of our simulations
the merging process with many minor mergers is not completed by
$T=10$\,Gyr.  While the results of our 1:1 to 1:10 mergers are
consistent with the growth observed in the simulations of
\cite{2012arXiv1206.1597H, 2012arXiv1206.5004H} and \cite{2012MNRAS.tmp...42O}, 
they are inconsistent with \cite{2009ApJ...703.1531N}.  The reason for this
discrepancy cannot be the steeper density slope of the stellar bulge,
as was suggested by \cite{2012arXiv1206.1597H}, because we tried even
higher density slopes for the minor child galaxies and found that this
does not affect the growth of the merger product (see
Appendix\,\ref{Sect:App:Density}).  The discrepancy between our
results and those of \cite{2009ApJ...703.1531N, 2009ApJ...706L..86N}
could be explained by the averaging of the simulation results, as was
also suggested by \cite{2012arXiv1206.1597H}.

\section{Constraining the model parameters}
\label{Sect:ConstrainingIC}

To perform the simulations we use an updated version of the
gravitational $N$-body tree-code {\tt Bonsai}
\citep{2012JCoPh.231.2825B}. All simulations are performed on
workstations with NVIDIA GTX480 graphical processing units.
Simulations are run with 1 to 18 GPUs, depending on $N$.

The code has three important parameters; the choice of the time step
$dt$, the softening length $\epsilon$ and the tree-code opening angle
$\theta$.  The opening angle is set to $\theta=0.5$ whereby we use the
minimum distance opening criterion \citep{1994JCoPh.111..136S}, which
is computationally more expensive than the improved Barnes-Hut method
\citep{1996NewA....1..133D} by about a factor of two, but offers better
accuracy \citep{2012JCoPh.231.2825B}.

After having fixed $\theta$, the choice of $dt$ and $\epsilon$ are
quite critical for the quality of our results.  We measure the quality
of the simulation based on the error in the energy and the change in
size of a galaxy model in isolation.  We determine the optimal values
for $dt$ and $\epsilon$ by performing an analysis in which we run a
series of models using a range of values for $dt$ and $\epsilon$. 
Although the time step and softening are discussed extensively in 
previous literature (e.g. \citep{1996AJ....111.2462M,
2000MNRAS.314..475A, 2001MNRAS.324..273D}) we chose to do extra tests 
our-self. This for a couple of reasons, first most of the previous discussed 
methods propose a softening that scales with the number of particles. 
However, we use different number of particles for the primary galaxy 
and child galaxies so we had to find a configuration that works for 
both. Second, our tree-code uses an accurate multipole expansion,
in combination with a opening angle criteria that is different from 
used in the previously mentioned papers. And finally we use two 
component models with power law density distributions while most 
examples in the previous mentioned papers are based on one component
spherical Plummer models. Therefore we decided to 
rather empirically test how the softening and time-step affected 
the properties of our initial conditions instead of taking over 
a previously result obtained in a different setting. 
The
initial conditions for these isolated galaxies are generated using
GalactICS
\citep{1995MNRAS.277.1341K,2005ApJ...631..838W,2008ApJ...679.1239W}
and consist of a dark matter halo and a baryonic bulge. All models
represent elliptical E0 galaxies. The bulges are modeled with a
Sersic index of 3 and the dark-matter halo is represented by an NFW
profile \citep{1996ApJ...462..563N}.

\begin{figure} 
 \includegraphics[width=1.0\columnwidth]{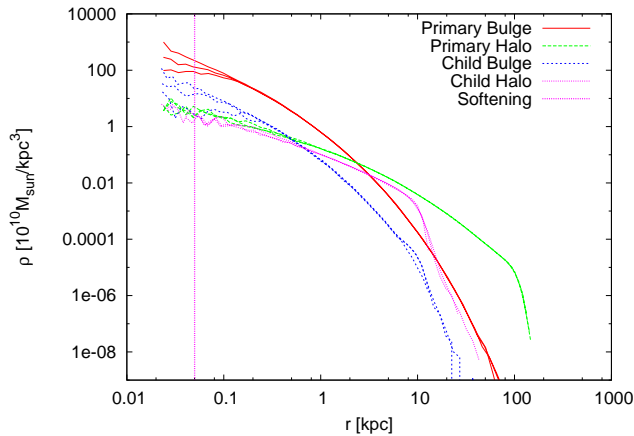}
 \caption{Density profiles of the HR baryonic bulge and dark-matter
   halos for the primary and child galaxies, which have evolved in
   isolation. For each component we present the density profile at
   zero age, at an age of 1\,Gyr and at 10\,Gyr. However, because the
   lines are almost indistinguishable and to prevent clutter we show
   each epoch with the same line style.  The expansion in time of the
   child halo is caused by the escape of a few dark-matter
   particles. This is also the main reason for the expansion of the
   selected model, noticeable in the right hand-panel of
   Fig.\,\ref{fig:ISOTimeSoftHRChild}.  The vertical line near
   $R=0.05$\,kpc indicates the softening length in the HR simulation.
 }
 \label{fig:ISODensity}        
\end{figure}

In our simulations we collide the primary galaxy with a number of
child galaxies, which are smaller and less massive.  The mass ratio
and the number of child galaxies which interact with the primary is
one of the free parameters in our simulations.  We generate the child
galaxies using GalactICS by adopting the appropriate mass and size
of the dark-matter halo both of which we assume to be a constant
factor smaller than in the primary galaxy. This factor is equal to the
mass ratio of the child, e.g. for a 1:10 child the mass and cut-off radius
is 10 times lower than that of the primary galaxy. As a consequence the density
of the child galaxy is smaller by a factor $\sim$10 than that of the primary. 
In Appendix\,\ref{Sect:App:Density} 
we demonstrate that the effect of our selected size and mass 
(and consequently density) of the child galaxy is
not critical for our conclusions.
In Fig.\,\ref{fig:ISODensity} we present the density profiles of the
baryonic bulge and dark matter halo of our selected initial conditions
at zero-age, at $T=1$\,Gyr which is the moment we intend to use them
for the merger simulations in \S\,\ref{Sect:InitConditions}, and at $T=10$\,Gyr. 
It takes a few crossing times for the
galaxies to relax after generation, therefore we first let them evolve 
for $T=1$\,Gyr before we use them in a merger configuration. The
differences in the density profile for the same component are hardly
noticeable as a function of time.   To prevent spurious mass
segregation all particles in one simulation have the same mass.  Mass
differences between galaxies and baryonic and dark matter particle
distributions are created by using different amounts of particles for
each galaxy and galaxy components (see Tab.~\ref{Tab:GalProgen} for
details).

The isolated galaxy simulations were performed for a 
galaxy with a total mass $M_{\rm parent} = 2.2 \times 10^{12}$\,\Msun,
and a half-mass radius of 1.36\,kpc.  which we call the
parent or primary galaxy and for each of the child galaxies.
The child galaxies have a mass that is $q$ times lower than
that of the primary. For example the $q=$1:10 child has a 
mass of $M_{\rm child} = 2.2 \times 10^{11}$\,\Msun.
To make sure that the child galaxies consist of enough particles 
to be accurately resolved and stay stable in isolation we use 
different resolution models for the initial conditions. Depending
on the number of child galaxies (mass ratio) used. We found that 
using a minimum number of $10^3$ ($10^4)$ particles for the bulge 
(dark matter) component 
proofed to be enough to keep the galaxies stable (Tab.~\ref{Tab:GalProgen}).

\begin{figure*}

 \includegraphics[width=2.0\columnwidth]{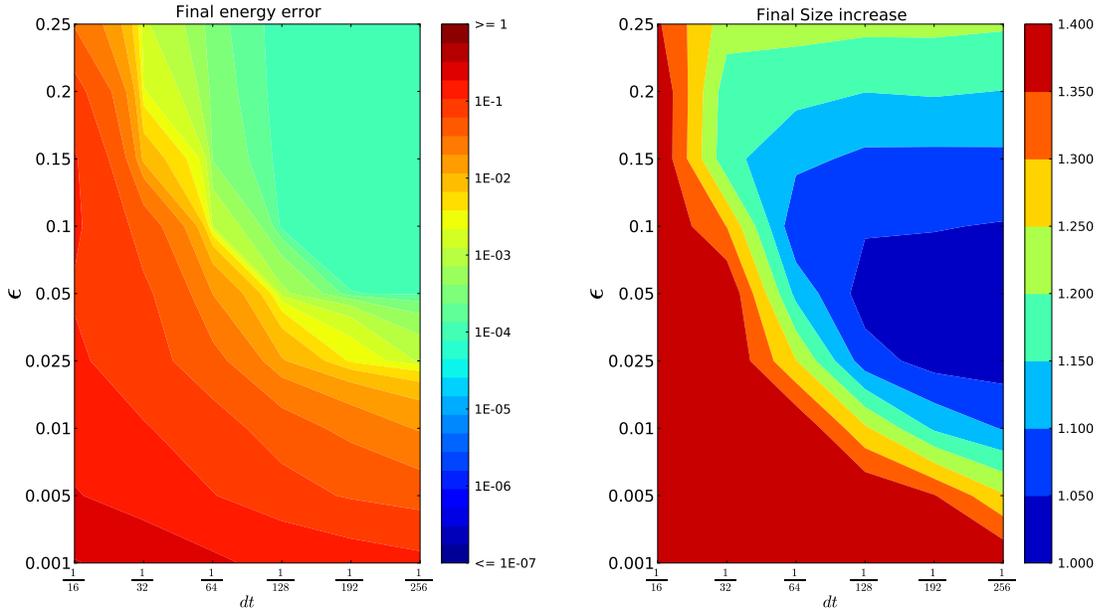}
 \caption{Results of the test simulations for the HR primary galaxies.
   The left panel gives the energy error $\Delta E$, the right panel
   gives the size increase {$\cal \gamma$} after 10\,Gyr (see
   Eq.\,\ref{Eq:expansionFactor}). Both panels present the information
   as a function of the time step ($dt$) and the softening length
   ($\epsilon$). The simulations are performed at the grid-points in $dt$
   and $\epsilon$, as indicated along the axes. The color scheme shows a
   linear interpolation between the grid points.  The
   scaling to the colors are provided along the right-hand side of
   each panel, and exhibits a log-scale for the left hand panel and a
   linear scaling for the right-hand panel.  }
 \label{fig:ISOTimeSoftHRMain}        
\end{figure*}

\begin{figure*}
 \includegraphics[width=2.0\columnwidth]{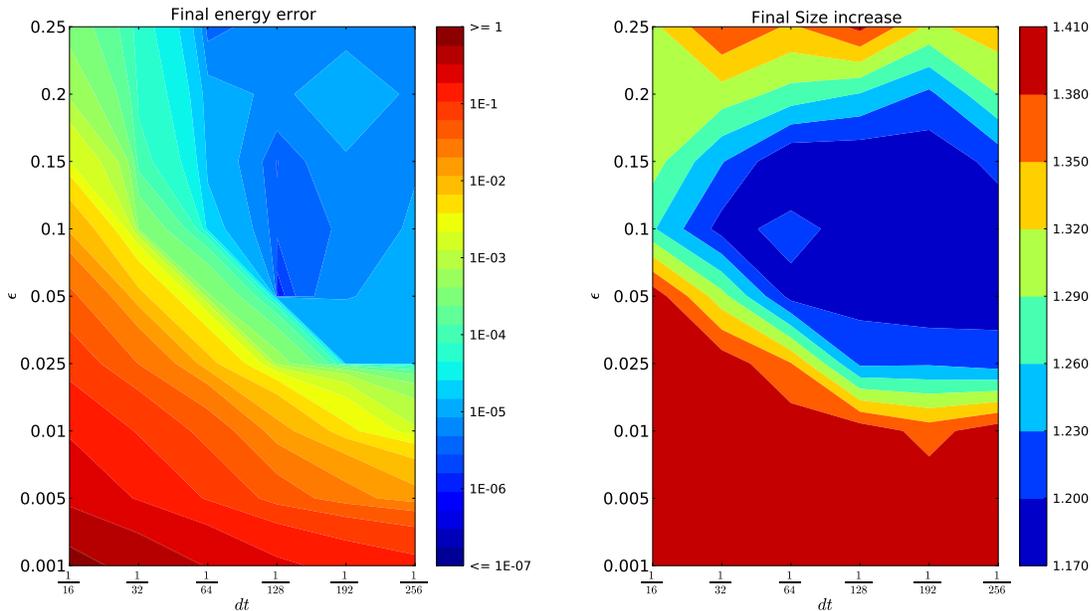}
 \caption{Results of the test simulations for the HR child galaxies,
   which are ten times less massive than the primary.  See
   Fig.~\ref{fig:ISOTimeSoftHRMain} for a further description of the
   figure.}
 \label{fig:ISOTimeSoftHRChild}        
\end{figure*}

We refer to low-resolution simulations (LR) for those in which the
primary galaxy was modeled with $N=2.2 \times 10^5$ and the 1:10 child with
$N=2.2 \times 10^4$. For the high-resolution (HR) simulations we
adopted $N=2.2 \times 10^6$ and $N=2.2 \times 10^5$ for the primary
and 1:10 child galaxies, respectively. The child galaxies with a different
mass ratio are scaled in a similar way as the 1:10 child,
namely the number of particles of the primary divided by the mass ratio.
Each model was run for 10\,Gyr.
In Figs.\,\ref{fig:ISOTimeSoftHRMain} and \ref{fig:ISOTimeSoftHRChild}
we present the results of the HR simulations; the error in the energy 
$\Delta E=(E_{10Gyr} - E_{0}) / E_{0}$ for the
left-hand panels and the expansion factor of the bulge, {$\cal \gamma$}(t),
in isolation in the right-hand panels.
\begin{equation}
   {\cal \gamma}(t) \equiv R_{h-t}/R_{h-0Gyr}
\end{equation}\label{Eq:expansionFactor}
Here $R_{h-t}$ is the half-mass radius of the baryonic component of
the galaxy at time $t$. In Fig.\,\ref{fig:ISOTimeSoftHRMain} we present the
result for the primaries, and
Fig.\,\ref{fig:ISOTimeSoftHRChild} for the 1:10 children.

As expected, $\Delta E$ decreases for
decreasing $dt$ and increasing $\epsilon$, for the primary as well as for
the child galaxy.  The growth in size of the galaxies becomes smaller
with decreasing $dt$.  However, for $\epsilon$ this is not a trivial
matter.  For too large a softening the galaxy tends to expand
substantially, but the effect is not as dramatic as for a too small
value of $\epsilon$.  The growth for large $\epsilon$ is caused by spurious
suppression of relaxation in the large $\epsilon$ runs.  The growth for
small $\epsilon$ is numerical, caused by strong encounters which are not resolved
accurately in our simulation. This limitation in our numerical solver
is also noticeable in the energy error for small $\epsilon$ 
(see Figs.\,\ref{fig:ISOTimeSoftHRMain} and Fig.\,\ref{fig:ISOTimeSoftHRChild}).

Because we intend to study the expansion of the primary galaxies caused by
mergers they should not expand in isolation. 
For our production simulations (see \S\,\ref{Sect:InitConditions}) we select the
$\epsilon$ which show the least expansion when the galaxy is run in
isolation. The time step is chosen as large as possible (for performance)
but giving the
smallest possible energy error.  We make the choice of $dt$ and $\epsilon$ based on the
simulations for the primary galaxy, which for the child also turns out
to be the bests choice.  For the HR simulations $dt=1/192$ and $\epsilon =
0.05$, and for the LR simulations $dt=1/128$ and $\epsilon = 0.1$.  Using
these settings the energy error of all the merger simulations stays
below $\ll 10^{-4}$.
These parameters are presented in dimension-less $N$-body units
\citep{1986LNP...267..233H}, but in physical units they translate to a
time step of $dt \simeq 0.078$\,Myr for the LR and $dt \simeq
0.052$\,Myr for the HR simulations, and a softening length of
$\epsilon\simeq 100$\,pc, and $\epsilon \simeq 50$\,pc for the LR and HR runs,
respectively.

Comparing the most optimal settings with the results of previous 
work on softening values \citep{1996AJ....111.2462M,
2000MNRAS.314..475A, 2001MNRAS.324..273D}. The chosen softening
values are comparable to the values advised in \citep{2000MNRAS.314..475A}
and \citep{2001MNRAS.324..273D}
with our chosen values being slightly larger to keep the energy
error less than $10^{-3}$ while the size growth stays smaller than 10\%. 
The difference between our results and \cite{1996AJ....111.2462M} 
is somewhat larger, but still within a factor of three. The difference 
can be attributed to the fact that we use more particles and use a 
tree-code instead of direct summation (see also Table 1.
of \cite{1998MNRAS.293..369A}).

Most of the expansion of our galaxies occurs in the first few
100\,Myr of the simulation, and therefore all galaxies are evolved
to an age of 1\,Gyr in isolation before they are used as initial
galaxies in the merger simulations.

As as extra test we ran several of the isolated galaxies using the
direct $N$-body code {\tt phiGRAPE} \citep{2007NewA...12..357H} to
confirm that the observed $\Delta E$ and expansion are not the result
of using an approximation based simulation code. The results of these 
simulations are consistent with the results of {\tt Bonsai}, but 
show a $\Delta E$  of $\sim10^{-6}$, because of the higher accuracy of {\tt phiGRAPE} .

With this choice of initial conditions, the child galaxies in the low
resolution simulations still expand by at most a factor of 2 before
the merger starts.
To check how this affects our results we redo the simulation but 
at the start the child galaxies are represented as point masses. 
The point mass is
replaced by a child galaxy as soon as it enters the dark matter halo
of the primary galaxy.
This simulation gives a nearly identical expansion of the primary
galaxy to that in which the child was resolved from the start of the
simulation.

\begin{table*}
    \begin{tabular}{ | l | c |  c | c | c | c | c | c | c | c | c | c | c | c | c |} \hline   
     \multicolumn{3}{|c|}{}          & \multicolumn{3}{|c|}{LR}   & \multicolumn{3}{|c|}{HR}\\ \hline
     \multirow{2}{*}{Model}          &  \multirow{2}{*}{Type}  &  Total mass & $N$    & P mass     & $\epsilon$ & $N$    & P mass     & $\epsilon$ \\ 
                                                       &       &  [$\Msun$]  & -    &  [$\Msun$] & [pc]       & -    &  [$\Msun$] & [pc]\\  \hline
     \multirow{2}{*}{Primary}                          &  Halo   &  $2\times10^{12}$  & $1\times10^5$   &  $1\times10^7$ & 100       & $2\times10^6$    &  $1\times10^6$ & 50  \\  
                                                       &  Bulge  &  $2\times10^{11}$  & $1\times10^4$   &  $1\times10^7$ & 100       & $2\times10^5$    &  $1\times10^6$ & 50  \\  \hline
     \multirow{2}{*}{Child $1\over5$}                  &  Halo   &  $4\times10^{11}$  & $4\times10^4$     &  $1\times10^7$ & 100       & $4\times10^5$    &  $1\times10^6$ & 50  \\  
                                                       &  Bulge  &  $4\times10^{10}$  & $4\times10^3$     &  $1\times10^7$ & 100       & $4\times10^4$    &  $1\times10^6$ & 50  \\  \hline
     \multirow{2}{*}{Child $1\over10$}                 &  Halo   &  $2\times10^{11}$  & $2\times10^4$     &  $1\times10^7$ & 100       & $2\times10^5$    &  $1\times10^6$ & 50 \\  
                                                       &  Bulge  &  $2\times10^{10}$  & $2\times10^3$     &  $1\times10^7$ & 100       & $2\times10^4$    &  $1\times10^6$ & 50 \\  \hline
     \multirow{2}{*}{Child $1\over20$}                 &  Halo   &  $1\times10^{11}$  &  -   &  - & -       & $1\times10^5$    &  $1\times10^6$ & 50 \\  
                                                       &  Bulge  &  $1\times10^{10}$  & -     &  - & -       & $1\times10^4$    &  $1\times10^6$ & 50 \\  \hline
     \multirow{2}{*}{Child $1\over40$}                 &  Halo   &  $5\times10^{10}$  &  -   &  - & -       & $5\times10^4$    &  $1\times10^6$ & 50 \\  
                                                       &  Bulge  &  $5\times10^{9}$  & -     &  - & -       & $5\times10^3$    &  $1\times10^6$ & 50 \\  \hline
     \multirow{2}{*}{Child $1\over80$}                 &  Halo   &  $2.5\times10^{10}$  &  -   &  - & -       & $2.5\times10^4$    &  $1\times10^6$ & 50 \\ 
                                                       &  Bulge  &  $2.5\times10^{9}$  & -     &  - & -       & $2.5\times10^3$    &  $1\times10^6$ & 50 \\  \hline

    \end{tabular} 

\caption{Galaxy properties. Characteristics of the base models used in the simulations. 
         The first column indicates if the galaxy is the primary galaxy or one of the minor merger
         child galaxies. The second column indicates if the properties are for the dark matter halo or for the 
         baryonic bulge and directly after in the third column the mass of the galaxy component. The next three
         columns show the properties when using low resolution models, the number of particles, mass per particle and the softening used.
         The final three columns show the number of particles, mass per particle and 
         softening used when using high resolution models.}
\label{Tab:GalProgen}
\end{table*}

\section{Initializing the galaxy mergers}
\label{Sect:InitConditions}

We use the isolated galaxies discussed in the previous \S\, to study
the effect of mergers. For this purpose we adopt the primary galaxy
and have it interact with an identical copy or with a number of
child galaxies. We refer to minor mergers when the primary:child mass
$\le$ 1:5. Major mergers in our study have equal mass.

\subsection{Configuring the major mergers}
\label{Sect:MajorMergerModels}

In the major merger simulations two identical copies of the primary
galaxy are placed on an elliptical orbit.  The initial distance
between the two galaxies is 400\,kpc, which exceeds the size of the
dark-matter halos, which is about 200\,kpc. The relative velocity is
chosen such that the minimal distance of approach during the first
perigalactic passage $P$ varies.  We vary $P$ from head-on (0\,kpc) to 10, 50,
75 and 100\,kpc for the widest encounters. All our merger simulations
are run until 10\,Gyr, after which both galaxies have merged to a
single galaxy, except for the $P=100$\,kpc configurations (see \S\
\ref{Sect:Results}).

Because we use two identical copies of the primary galaxy the galaxies 
will be co-rotating during the collision. To study the influence of the rotation angle
on the size of the merger product we run the simulations for $P=0$ and
$P=50$\,kpc with co-rotating and counter-rotating galaxies. This is 
done by changing the inclination and $\Omega$ of one of the two galaxies
between 0\,$^{\circ}$ and 270\,$^{\circ}$ in steps of 90\,$^{\circ}$ using
a rotation matrix. After the rotation the galaxies are put on their
elliptical orbit as described in the previous \S. The stability of the isolated 
galaxies is not affected, because the positions and velocities are rotated
in a consistent way. We did not perform this study on the other major
mergers, because the effect of the inclination and $\Omega$ 
on the expansion factor turns out to be negligible (variation
of the final size between the 16 configurations is $\sim 0.02$\,kpc).

\subsection{Configuring the minor mergers}
\label{Sect:PlummerModels}

The easiest way to generate an initial configuration with 1 primary
galaxy and $N$ child galaxies is by randomly positioning them in a
spherical distribution. For each simulation we generate a
Plummer distribution~\citep{1915MNRAS..76..107P} of $N+1$ particles.
We then select a random particle and reposition it in the center of
mass. We assign the mass of the parent galaxy to this central
particle.  The other particles are assigned the mass of the child
galaxies.  The total mass of all child galaxies in each simulation is
the same as that of the primary galaxy.  The model is then rescaled to
a virial radius, $R$, and virial ratio (temperature), $Q$. We run 10
sets of initial conditions for each selected mass ratio.  In the next
section we discuss the results of a number of simulations with
varying mass-ratio, $R$ and $Q$.  Using this procedure we created a
range of minor merger models with varying $R$, $Q=0$ and $N$, see
Tab.\,\ref{Tab:ICs}.  In Section~\ref{Sect:VaryTemp} we relax
this assumption and allow $Q>0$ to study the effect of the
temperature on the size of the merger product.

\begin{table}
\begin{tabular}{lllll}
\hline
mass ratio & R   & Q & resolution & name \\
\hline
  1:5      & 500 & 0 & LR & 1:5\\
  1:10     & 200 & 0 & LR & 1:10, R200\\
  1:10\footnotemark[1]
           & 500 & 0 & LR & 1:10, R500 \\
  1:20     & 500 & 0 & HR & 1:20\\
  1:40     & 500 & 0 & HR & 1:40\\
  1:80     & 500 & 0 & HR & 1:80\\
\hline
  1:10     & 200 & 0.0 & LR & \\
  1:10     & 200 & 0.05 & LR &  \\
  1:10     & 200 & 0.1 & LR & \\
  1:10     & 200 & 0.3 & LR & \\
  1:10     & 200 & 0.5 & LR & \\
  1:10     & 200 & 1.0 & LR & \\
  1:10     & 500 & 0.0 & LR & \\
  1:10     & 500 & 0.05 & LR & \\
  1:10     & 500 & 0.1 & LR & \\
  1:10     & 500 & 0.3 & LR & \\
  1:10     & 500 & 0.5 & LR & \\
  1:10     & 500 & 1.0 & LR & \\
\hline
\end{tabular}
\caption{Initial conditions for the minor merger simulations. The first column indicates 
the used mass ratio. The second and third column indicate the virial radius, $R$, and virial temperature, $Q$,
 to which the configuration is scaled. The fourth column indicates the used resolution, either 
low-resolution ($N=4.4 \times 10^5$) or 
high-resolution ($N=4.4 \times 10^6$). The final column indicates the name used in the text
when we refer to the configuration. }
\label{Tab:ICs}
\end{table}
\footnotetext[1]{based on same Plummer distribution as the $R=200$ models.}
\addtocounter{footnote}{1}

For the 1:10 minor mergers we adopted $R=200$\,kpc as well as
$R=500$\,kpc.  In the $R=200$ configurations the child galaxies have a
$\sim$\,50\% probability of being placed initially within the dark
matter halo of the primary. For the $R=500$ configurations, which are
based on the same random Plummer realization but scaled to a larger
virial radius, this is only $\sim$\,15\%.

\subsubsection{Plummer models - Varying virial temperatures}
\label{Sect:VaryTemp}

In order to study the effect of $Q$ on our results we selected one of
the 1:10 models. The merger remnant of the selected model has
roughly the same mass in both the $R=200$ and $R=500$
configurations indicating that they were involved in an equal number of
mergers by the end of the simulation.  We used this initial
configuration to vary $Q$ between 0 (cold) and 1 (warm), see
Tab.\,\ref{Tab:ICs}. Each run was repeated with $R=200$\,kpc and for
$R=500$\,kpc.  This results in a total of 12 different configurations,
all of which are based on the same spatial distribution but with
different virial radii and virial temperatures.

\section{Results}\label{Sect:Results}

\begin{figure}
 \includegraphics[width=0.7\columnwidth, angle=-90]{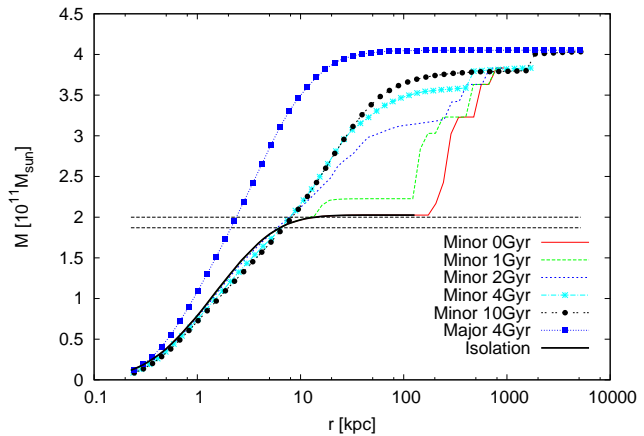}
 \caption{Cumulative mass profiles of the bulge. Shown are the profiles of one of the 1:10
          merger simulations with $R=500$ and $Q=0$ and the major merger simulation with
          $P=50$\,kpc. The thick solid line shows the cumulative mass profile of the 
          main galaxy at the start of the simulation. The horizontal lines show the half-mass 
          of the major merger (top line at $2.0M_{\sun}^{11}$) and of the minor merger
          (bottom line at $1.86M_{\sun}^{11}$). The remaining lines show cumulative mass profiles
          of the minor merger as indicated.}
 \label{fig:PlumSLExample1}        
\end{figure}

During the simulation snapshots are taken every 100M year and
post-processed using {\tt
  tipsy}~\footnote{http://www-hpcc.astro.washington.edu/tools/tipsy/tipsy.html}.
We compute the density and cumulative mass profiles which are
subsequently used to determine the half-mass radius ($R_h$), 
measured using the baryonic particles only.
To compute the profiles {\tt tipsy} computes the densest point in the
simulation and then uses spherical shells centered on the densest
point to bin the particles.  To compute $R_h$ we can not simply take
the radius that contains half of the total mass in the merger system,
because depending on the time of the snapshot, not all child galaxies
have merged with the primary galaxy.  Therefore we analyze the
cumulative mass profile (based on all baryonic particles in the system) 
of the galaxy and take as total mass the point
where the profile flattens.  This flattening indicates that all child
galaxies within that radius have merged\footnote{We chose this method 
over a method where we check if particles are bound, since in this 
method we include child galaxies that would add to the luminosity of 
the remnant galaxy if they would be observed.} with the primary galaxy and
that the unmerged child galaxies are too far away to be counted as
part of the merger remnant. This total mass is then used to determine
$R_{h}$. We illustrate this procedure in
Fig.~\ref{fig:PlumSLExample1}.

In Fig.~\ref{fig:PlumSLExample1} we present the cumulative mass-profiles
of one of the 1:10 mergers with $R=500$ and $Q=0$, for the major
merger with $P=50$\,kpc and the profile of the isolated
primary galaxy.  During the first four Gyr of evolution the cumulative
mass of the minor merger simulation gradually increases each time one
of the child galaxies is accreted.
We determine the total mass of the merger remnant
by determining the first moment that the cumulative mass profile
becomes flat.  This happens, for example, for the 1\,Gyr situation at
a mass of $M\sim2.2 \times 10^{11} M_{\sun}$, and for 2\,Gyr at
$M\sim3.2 \times 10^{11}M_{\sun}$.
We further notice that the minor merger remnant has grown in size with 
respect to the major merger remnant, even though it has accreted less mass. 
The major merger remnant ends up with twice the mass of the primary galaxy,
e.g. a complete merger of all mass in the system. Whereas in the minor merger
only 8 child galaxies have been accreted in the 4\,Gyr after the start of the simulation. At the end of the 
simulation (T=10\,Gyr) 9 child galaxies have merged with the primary galaxy. The last unmerged 
child appears in the cumulative mass distribution (striped line with solid circles),
represented by the bump in the line at $r\sim1800$\,kpc (Fig.~\ref{fig:PlumSLExample1}).
As a consequence the total mass of the merger product at the end of the simulation 
is larger for the major merger than for the minor merger.

\subsection{The growth of the primary due to subsequent mergers}
\label{Sect:NumberOfMergers}

\begin{figure}
 \includegraphics[width=0.7\columnwidth, angle=-90]{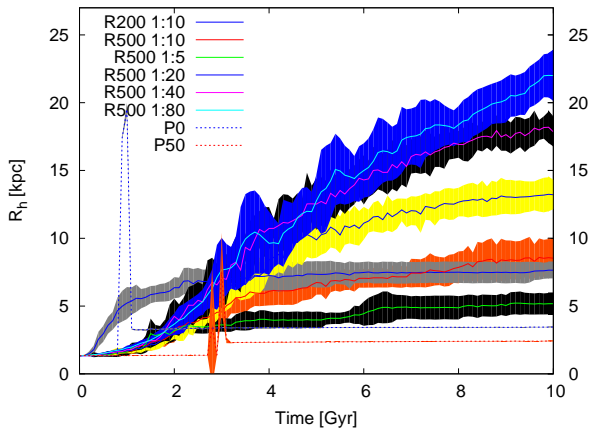}
 \caption{Evolution of $R_{h}$. The plot shows the mean (solid and
   dotted lines) and the standard deviation (shaded areas) of the
   $R_{h}$ based on different merger configurations. The solid lines
   show the results of the minor merger configurations.  The dotted
   lines show the major merger configurations with $P=0$ and $P=50$.
   The minor mergers are each averaged over 10 different Plummer
   realizations, the major mergers are averaged over 16 different
   combinations of Euler angles. For 1:80 we filtered out size
   jumps caused by invalid size detection.  }
 \label{fig:bulkEvolQ0Radius}        
\end{figure}

\begin{figure}
 \includegraphics[width=0.7\columnwidth, angle=-90]{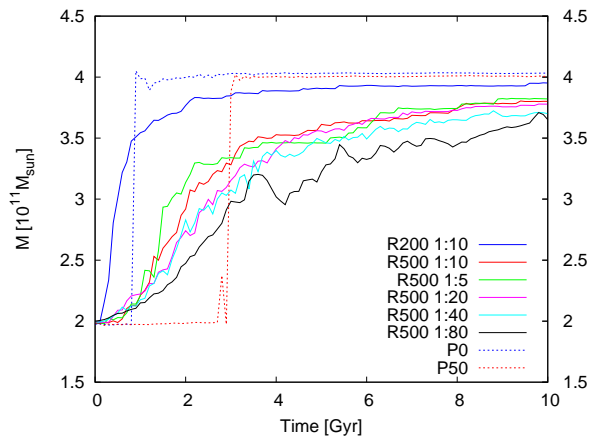}
 \caption{Evolution of the merger products mass. The same simulations are 
          shown as in Fig.~\ref{fig:bulkEvolQ0Radius}, but instead of the radius 
          the average mass of each set of simulations is shown.}
 \label{fig:bulkEvolQ0Mass}        
\end{figure}

We run each major merger model 16 times (with different inclination and  $\Omega$ ) and average the $R_{h}$.
For each set of minor merger initial conditions ($R$ and $N$, see Tab.\,\ref{Tab:ICs}.) 
we perform 10 simulations and as in the major mergers we average $R_{h}$.
These results are presented in Fig.~\ref{fig:bulkEvolQ0Radius} for
the growth in size, and Fig.~\ref{fig:bulkEvolQ0Mass} for the
growth in mass. The shaded areas are the $1\sigma$ deviation from the mean,
but for the major mergers the areas are barely visible in comparison with the width of the line. This
indicates that the inclination and the angle $\Omega$ hardly affects the
growth of the merger product.  Only during the merger event itself
some deviation is noticeable (at about 3\,Gyr for the simulation with
$P=50$\,kpc). The head-on configuration ($P=0$) grows from an initial
size of $R_h=1.3$\, kpc to 3.4\,kpc, whereas the $P=10$ to 75\,kpc
simulations grow to only about 2.4\,kpc (only the $P=50$\,kpc is presented
in Fig.~\ref{fig:bulkEvolQ0Radius}). The $P=100$\,kpc did not
experience a merger in our simulation within the time frame of 10\,Gyr.
For the minor merger configurations the result is quite different as can be 
seen in Fig.~\ref{fig:bulkEvolQ0Radius}. The shaded areas indicate
that there is variation in size between the different configurations.
The size growth however is larger than that of the major mergers. 
The results indicate that the growth in size is directly
related to the mass ratio of the child galaxies, the larger the difference
in mass the larger the growth in size. For the 1:10 simulations the $R_h$ is 
at least 7\,kpc and for the 1:80 simulations the average $R_h$ is at least 20\,kpc 
while at the start of the simulation the primary galaxy has a $R_h$ of 1.3\,kpc.

To study the effect of increasing the number of minor mergers  (and
decreasing the mass per child galaxy)
in more detail we start by comparing the major
merger with the 1:5 minor merger. This case results in a $R_h$ of
5.1\,kpc, averaged over 10 simulations. The $1\sigma$ deviation 
indicates that the result is rather consistent when rerunning with a different initial
Plummer realization. The extreme values of $R_h$ after 10\,Gyr range between 4 and
6\,kpc. Even the increase due to the 1:5 minor merger exceeds the
$P=0$ major merger case by a factor of $\sim 1.2$ to 1.8.  By the end
of the simulation at 10\,Gyr on average 4 out of the 5 minor galaxies
have merged with the primary galaxy, and we expect that if the last
child galaxy would be accreted the increase in size would be even
larger.
The 1:10, R200\,kpc minor mergers show an average size of 7.6\,kpc, 
and 8.5\,kpc for the 1:10, R500\,kpc models; an increase of 5.8 to
6.5. The $R=200$\,kpc simulations merge on a shorter time scale compared to
the $R=500$\,kpc, which results in a smaller expansion, but higher
mass of the simulations performed in a smaller volume. This is also
visible in Fig.~\ref{fig:bulkEvolQ0Mass} where the $R=200$\,kpc increases
faster in mass and ends up being more massive than the $R=500$\,kpc case. 
Even though $R=500$\,kpc results in a less massive merger remnant than the
$R=200$\,kpc simulations, it grows to a larger size. This is caused by
the disturbing effect that the dark-matter halo of the primary galaxy
has on the child galaxies. With more galaxies placed outside the halo for 
the $R=500$\,kpc case this effect will be more pronounced.
While we continue to increase the number of child galaxies
the size of the merger remnant continues to grow. 
We stopped with $N=80$ children, at which moment the increase in
size compared to the major merger exceeded a factor 17.  The main
reason for the growth of the merger remnant when increasing the number
of child galaxies is by the heating of the merger remnant.  For
each doubling of the number of child galaxies from 1:5 to 1:10, etc.,
the growth in size of the merger product is about constant by $\sim
5$\,kpc.  The growth in mass for each of these initial configurations
is comparable, except for the 1:80, which grows less quickly.
The smaller mass growth in the 1:80 is attributed to increased 
dynamical interactions among the child galaxies. These interactions
cause the merging times per galaxy to take longer than when they would
merge sequentially.

\begin{figure}
 \includegraphics[width=0.7\columnwidth, angle=-90]{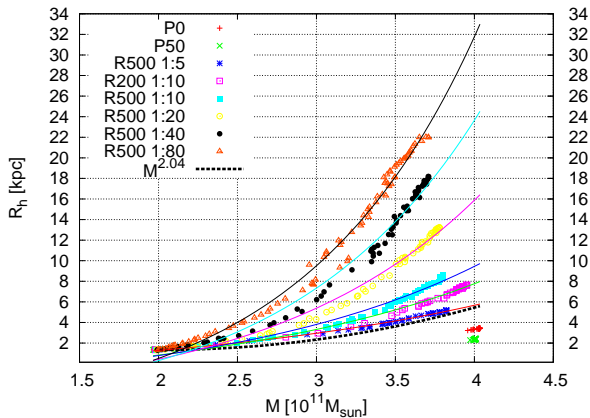}
 \caption{The mass of the merger remnant, $M$, versus $R_{h}$ for the different
   configurations. The solid lines are exponential fits through the results of the 
   minor merger configurations. The major mergers are not fitted. 
   The thick dashed line (bottom line) is the observed scaling relation of  
  $R_h \propto M^{2.04}$~\citep{2010ApJ...709.1018V}. Configurations are 
   plotted after each merger event (but filtered to not show fly-by events) and therefore 
   occur multiple times. The exponents of the fits are (from top to bottom), 0.9, 
   0.84, 0.74, 0.62, 0.57 and 0.48. If we make an exponential fit to the observed scaling
   relation than this would have an exponent of 0.5.} 
 \label{fig:MassVsRadBulk}        
\end{figure}

The combined growth in size and mass indicates an other effect, namely 
the larger the mass ratio the higher the growth in size. This is 
indicated in Fig.~\ref{fig:MassVsRadBulk} where we present $R_h$ 
as a function of the merger remnant mass (which is a function of time).  
This figure quantifies the growth in size of the merger remnant while 
increasing the number of mergers. Note that the mass ratio between the 
merger remnant and the child galaxies changes over time. For example the
1:10 minor mergers start with a mass difference of a factor 10 between
one child galaxy and the primary galaxy, but after the primary galaxy has 
merged with 5 child galaxies the mass ratio between the merger remnant 
and one child galaxy is 1:15. Interestingly it is not necessarily the mass
that drives the merger remnant size, but it is the number of children
that merges with the primary galaxy.  The solid curves are fits
through the data, and represent exponential curves. The dashed line 
is the observed scaling relation of  $R_h \propto M^{2.04}$~\citep{2010ApJ...709.1018V}. As we can see
our minor mergers are above this relation while the major mergers
are clearly smaller than the observational results. Indicating that
the minor mergers can cause a size increase comparable or even larger
than that observed.

\subsection{The effect on the galaxy shape due to subsequent mergers}

Now that we have established that the minor mergers can have a pronounced 
effect on the size of the galaxy remnant it will be interesting
to see if we can recognize this galaxy growth by the shape of the 
merger remnant. To test the effect that the mergers have on the shape
we measure the semi-principal axes $a$, $b$ and $c$ of the ellipsoidal 
with $a >= b >= c$. The axes lengths are obtained by computing the 
eigenvalues of the inertial tensor of the galaxy.
Using this we plot the three axis ratios $b/a$, $c/a$ and $b/c$, if these 
three ratios are 1 then the elliptical is perfectly spherical. 
If one of the axis is much smaller than the other two the galaxy 
is a flattened spherical. 

\begin{figure*}
 \includegraphics[width=1.7\columnwidth, angle=0]{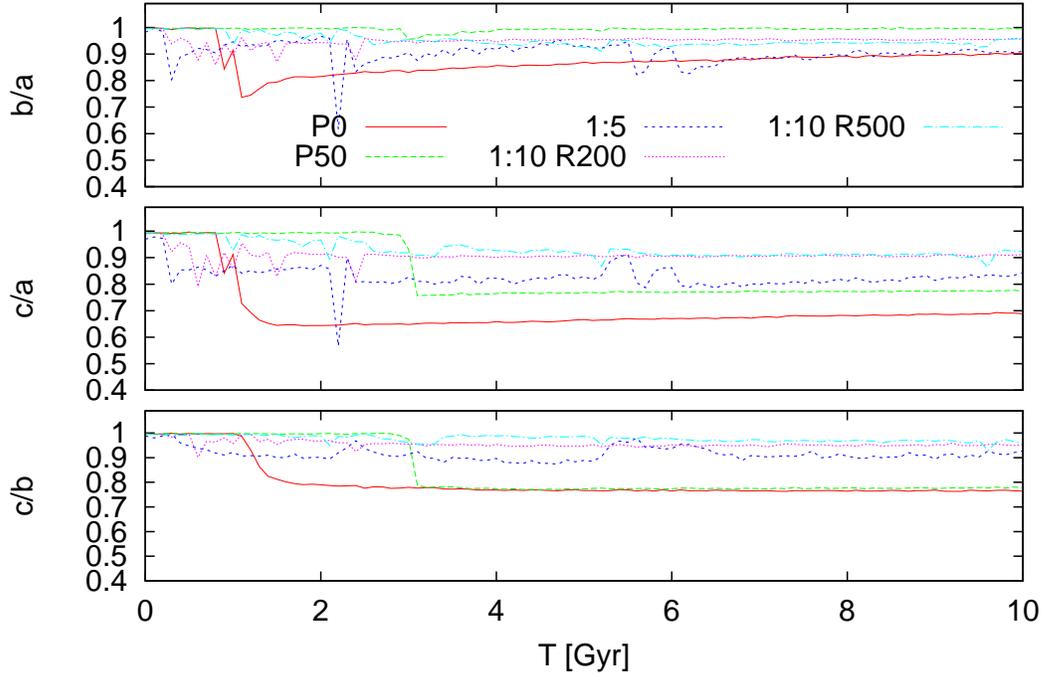}
 \caption{Axis ratios of the merger remnant. The three panels present the axis ratios 
    b/a, c/a and b/c of the merger remnant (top to bottom, with a $\geq$ b $\geq$ c). 
    As location to measure the shape is taken the mean $R_h$ of the configuration. If 
    all axis ratios are equal to 1 the shape is perfectly spherical. Shown are the results
    for the head-on major merger ($P=0$\,kpc), for the major merger with $P=50$\,kpc, 
    for a random 1:5 merger and for a random 1:10 merger once with $R=200$\,kpc
    and once with $R=500$\,kpc. 
  }
 \label{fig:shapeOverYear1}        
\end{figure*}

\begin{figure*}
 \includegraphics[width=1.7\columnwidth, angle=0]{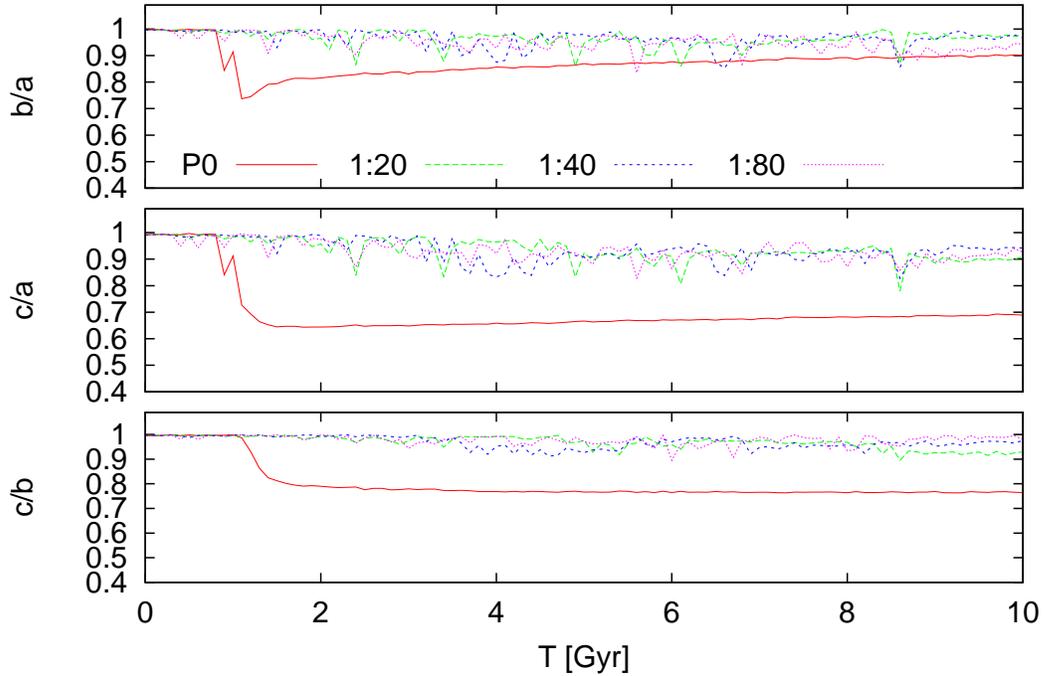}
 \caption{As Fig.~\ref{fig:shapeOverYear1}, but now for one of the 1:20, 1:40 and 1:80 configurations. 
          The head-on major merger from Fig.~\ref{fig:shapeOverYear1} is also presented (solid line).}
 \label{fig:shapeOverYear2}        
\end{figure*}

In Figs.~\ref{fig:shapeOverYear1} and ~\ref{fig:shapeOverYear2}
we present how the shape of the merger remnant (measured at $R_h$) 
evolves over time (baryonic component only). For the major mergers 
we used the default configurations 
with $P=0$ and 50\,kpc and for the minor merger configurations we 
randomly took one of the 10 realisations for each of the used mass ratios.
The x-axis shows the simulation time while each of the three panels presents one of 
the aforementioned axis ratios. 
Visible is that all galaxies start of as a perfect sphere, which is the 
initial condition, but over time the merger remnants deform. The major mergers 
show the highest degree of deformation with large differences between the 
different axis. The effect is most severe for the heads-on collision which is
expected as it is the most violent collision of all our configurations.
Although the major merger remnants become slightly more spherical after 
the mergers are complete the galaxies will not become perfectly spherical within a Hubble time. 

For 1:5 we notice, as with the the major mergers, a clear deformation from 
spherical, but the effect is less strong than for the major mergers. The individual
merger events of 1:5 are visible by the spikes in the axis ratios. The impact 
that a merger event has on the shape of the merger remnant becomes 
smaller when the mass difference between merger remnant and the infalling 
galaxy increases. 
For the 1:10 configurations, we do see slight deformations during the merger events 
(the shark tooth's in the lines), but when the merger is complete the remnant settles 
back to spherical. The same effect is visible for 1:20, 1:40 and 1:80 
(Fig.~\ref{fig:shapeOverYear2}). The final merger remnant of the 
minor merger configurations is no longer perfectly spherical, but it clearly is 
not as flat as the major merger configurations. Indicating that the accreted mass 
is distributed differently through out the merger remnant in the minor merger simulations 
than in the major merger simulations. We cover the location of the 
accreted mass in more detail in Section~\ref{Sect:Discussion}.

\begin{figure*}
 \includegraphics[width=1.7\columnwidth, angle=0]{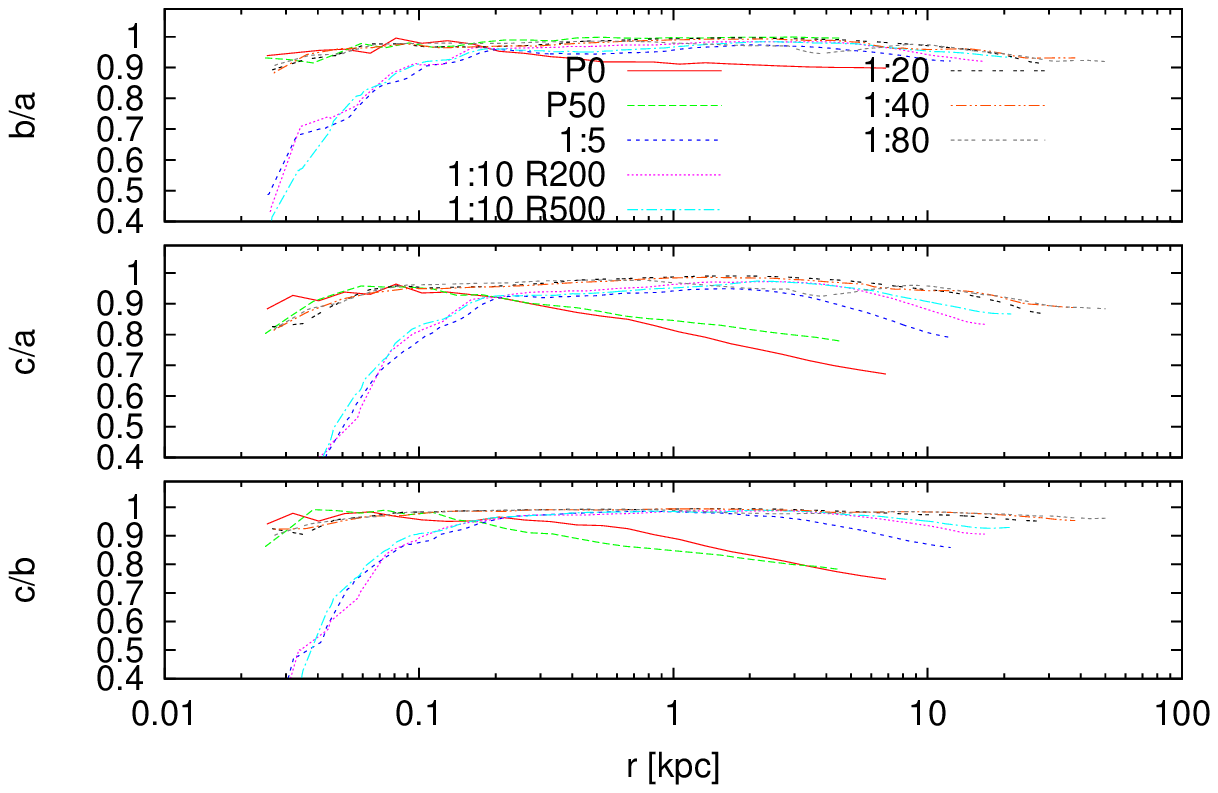}
 \caption{Final shape of merger remnant. On the x-axis we show the distance
  from the core of the merger remnant. The three panels present the axis ratios 
    b/a, c/a and b/c (top to bottom, with a $\geq$ b $\geq$ c). As final measurement 
    location we took $2 \times R_h$. The results for the minor mergers are averaged over all 10 
    different Plummer realisations. The differences in the cores of the 1:5 and 1:10 
    mergers is caused by the lower resolution with which these models have been run 
    (see text for details).}
 \label{fig:shapeOverRad}        
\end{figure*}

In Fig.~\ref{fig:shapeOverRad} we present the galaxy shape at $T=10$\,Gyr
from the core of the merger remnant up to $2\times R_h$, while in the 
previous paragraph we looked at the shape of the merger over time and only at the 
$R_h$ location. Because $R_h$ is different for each configuration
the lines representing the configurations in Fig.~\ref{fig:shapeOverRad} have different lengths.
The minor merger results are averaged over the different realisations of the model. 
We notice that the larger the differences in mass between the the primary galaxy
and the child galaxies the more spherical the merger remnant stays (axis ratios near 1).
The differences between the major and minor mergers are especially visible on the 
outsides of the merger remnant ($r > R_h$).
The major mergers are much more deformed from perfectly spherical into flattened 
ellipticals while the minor mergers stay near spherical all the way to the outskirts 
of the galaxy. The difference in the core of 1:5 and 1:10 with 
respect to the other configurations is the result of running these simulations with
only $N=4.4 \times 10^5$  particles and accompanying larger softening. We checked 
this by comparing the individual results of one of the major mergers and a 1:10 high 
resolution simulation with the comparable configurations in low resolution. In the high 
resolution simulations we do not observe the deformation in the core, instead
the axis-ratios have a similar profile as the other high resolution configurations.

\subsection{The effect of the virial temperature}

\begin{figure}
 \includegraphics[width=0.7\columnwidth, angle=-90]{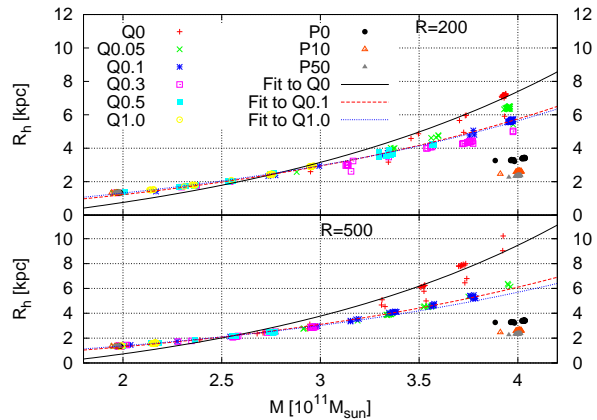}
 \caption{The mass, $M$, of the merger remnant versus the half-mass radius, $R_{h}$, for major merger
      configurations and a 1:10 merger configuration with different virial temperatures ($Q$).
      The minor mergers are all based on the same Plummer distribution, but rescaled to 
      a virial radius of $R=200$ ($R=500$) in the top (bottom) panel and a virial
      temperature which is varied between 0 (cold) and 1 (warm). Both panels show
      the major mergers with $P=$ 0, 10 and 50\,kpc.
      The lines are exponential fits through the configurations with $Q=0$ (solid), $Q=0.1$ (striped)
      and $Q=1.0$ (dotted). With exponents 0.57, 0.49 and 0.485 for $R=200$ and 
      0.62, 0.49 and 0.484 for $R=500$. Configurations are plotted 
      after each merger event and therefore occur multiple times.}
 \label{fig:MassvsRadiusColdToWarm}        
\end{figure}

In the previous section the virial temperature ($Q$) of the minor merger 
configurations was always set to zero. This resulted in cold collapse and 
consequently the fastest possible merger. In this section we investigate
the configurations described in 
Section~\ref{Sect:VaryTemp}  where $Q$ is varied. The results are presented in 
Fig.~\ref{fig:MassvsRadiusColdToWarm} where we present $R_h$ 
as a function of the merger remnant mass\footnote{We filtered the results to not include fly-by 
events which temporarily increases $R_h$.} (which is a function of time).
In the top panel the results for the $R=200$\,kpc are presented 
and the bottom panel presents the results for the $R=500$\,kpc
configurations. The results for the major mergers with $P=$ 0, 10 
and 50\,kpc are presented in both panels.

The major mergers are the same as the ones 
presented in the previous section and end up with a final
mass that is double that of the isolated galaxy
($M=\pm4\times10^{11}M_{\sun}$). Indicating that in all major 
merger configurations the two galaxies fully merge. The  
horizontal distribution in mass is caused by accretion of stars 
that were flung out and later fell back onto the merger product,
thereby increasing the mass of the merger remnant. 

The merger events that occur in the minor merger simulations can be seen
in the horizontal distribution of the points. The horizontal distance between
the points compares to a separation equal to the mass of a 1:10 child galaxy.
Even though the minor merger configurations are based on the same spatial distribution,
they are not per se involved in the same amount of mergers, this depends
on $R$ and $Q$. Comparing the cold ($Q \leq 0.1$) and warm ($Q > 0.1$) configurations we notice
that at the end of the simulation the cold configurations have a higher mass 
than the warm case. In the warm configurations the child galaxies 
have a higher velocity which increases their merging times. However, 
both warm and cold show a size increase larger than that of 
the major merger simulations. This is confirmed by the lines that show 
exponential fits\footnote{Note that the
fits through the $Q=1.0$ results are based on only 4 (1) merger events
in the $R=200$\,kpc ($R=500$\,kpc) configurations. The observed trend however is
the same as for the $Q=0.1$ configuration which is based on 10 (9) mergers.}
through the configurations with $Q=0$, $Q=0.1$ and $Q=1.0$.

The fits indicate that if all child galaxies merge with the primary,
the merger remnant would have a size that is at least a factor of two
larger than the remnants formed by major merging independent of $Q$.
For the $Q=0$ mergers this is even three times as large as the major mergers.
Also visible in Fig.~\ref{fig:MassvsRadiusColdToWarm} is the 
difference between the $R=200$\,kpc and $R=500$\,kpc configurations.
Here we notice, as in Fig.~\ref{fig:bulkEvolQ0Radius}, that the $R=500$\,kpc 
configurations show a larger size increase than the $R=200$\,kpc configurations, even though
the amount of completed mergers in the $R=500$\,kpc is less 
than in the matching $R=200$\,kpc configuration. This is attributed by
the higher energy input that the child galaxies give in the $R=500$\,kpc 
configuration compared to the $R=200$\,kpc situation where half the 
child galaxies are already within the dark matter halo of the primary
galaxy at $T=0$.

\section{Discussion}
\label{Sect:Discussion}

The simulations indicate that the effects of the minor mergers becomes
more pronounced when more mergers are involved in the mass growth.  To
test if this trend continues we performed three extra simulations with
160 child galaxies (mass ratio 1:160). With this many children the
simulation approaches the regime of a continuous stream of infalling
material.  The three simulations gave very similar results with little
variation between the runs.  With this many child galaxies their
distribution around the primary galaxy becomes almost uniform. The
effect of the random placement of the child galaxies is therefore much
smaller than in the simulations for 5 to 20 child galaxies.
Furthermore some of the children will merge with other children before
they reach the primary galaxy. For the 1:160 simulations we use $N=8.8
\times 10^6$ particles evenly divided over the primary ($N=4.4 \times
10^6$) and the child galaxies ($N=27500$ per galaxy).

\subsection{Properties of the merger remnant}

\begin{figure}
 \includegraphics[width=0.7\columnwidth, angle=-90]{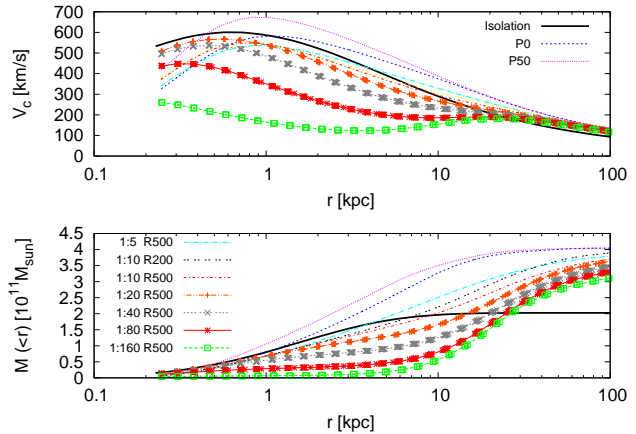}
 \caption{Circular velocity (top) and cumulative mass (bottom) profiles after 10\,Gyr for
   different merger configurations. Each line shows
   the averaged result of all random realisations for each of the
   configurations. The velocity and mass profile of the isolated model
   are indicated with the thick solid line.  }
 \label{fig:circVelocity}        
\end{figure}

The circular velocity ($V_c= \sqrt{M( < r)/r}$ ) 
and the cumulative mass profile at the end of the simulations (at $T=10$\,Gyr), both
are presented in Fig.~\ref{fig:circVelocity}. The top panel shows $V_c$ 
and the bottom panel presents the cumulative mass profile over the same distance
from the core. For the mergers we averaged the results of the random realizations.

The major mergers cause an increase in the $V_c$ compared to
the isolated model. Whereby the remnants formed by
major mergers with an elliptic orbit ($P=50$\,kpc) have a $V_c$
which is $\sim$100\,km/s higher than that of the isolated galaxy.  
For the remnants that formed by minor mergers we see a decrease in
$V_c$ compared to the isolated galaxy. 
Especially for the 1:40 and 1:80 the effect is quite pronounced, 
with the large size increase comes a decrease in velocity.  
This indicates that the lighter the child galaxies are the lower the 
$V_c$ becomes, this trend continuous for the 1:160 models. These models 
have the lowest circular velocity of all the configurations while
at the same time showing the largest size increase. 

\begin{figure}
 \includegraphics[width=1\columnwidth, angle=0]{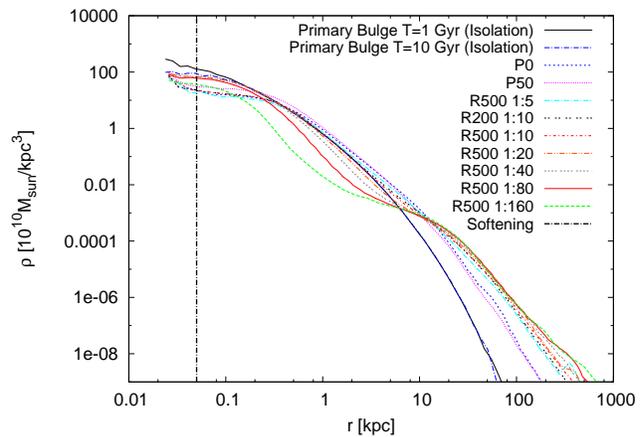}
 \caption{Density profiles of the different major and minor configurations 
          at the end of the simulation ($T=10$\,Gyr). Also displayed is the 
          initial density profile of the isolated galaxy. The vertical line
          is the softening used in the simulations.}
 \label{fig:FinalDensity}        
\end{figure}

The size increase can be deduced from the cumulative mass profiles
presented in the bottom panel of Fig.~\ref{fig:circVelocity}. We
notice that the mass in the core of the merger decreases when more
child galaxies are involved in the merging process.  This effect is
especially pronounced for the smaller mass-ratios ($<$1:20). The
smaller child galaxies have longer merging times, this combined with
their interactions inside the core of the merger remnant does not
allow the merger remnant to stabilize. Therefore the mass does not
sink back to the core within the simulation period.  This is also
apparent in the density profiles presented in
Fig.~\ref{fig:FinalDensity}.  Here the major mergers have a density
comparable to that of the isolated galaxy which in turn is higher than
the density of the minor mergers.  For the 1:5 and 1:10 mergers the
density profiles have a similar shape to that of the isolated galaxy
and the major merger remnants, but the configurations with mass ratios
of $\leq 1:20$ have a different shape.  The relatively sudden
depression in the density profiles of 1:80 and 1:160 near $r \sim
1$\,kpc is a trend that is visible in the 1:40 configuration and in
slightly less pronounced form for the 1:20 configurations.  The longer
merging times and dynamic interactions that takes place when the
number of child galaxies is $\geq$ 20 causes a slowdown in circular
velocity, a reduction in mass and therefore a drop in density in the
inner parts of the merger remnant.

\begin{figure}
 \includegraphics[width=0.7\columnwidth, angle=-90]{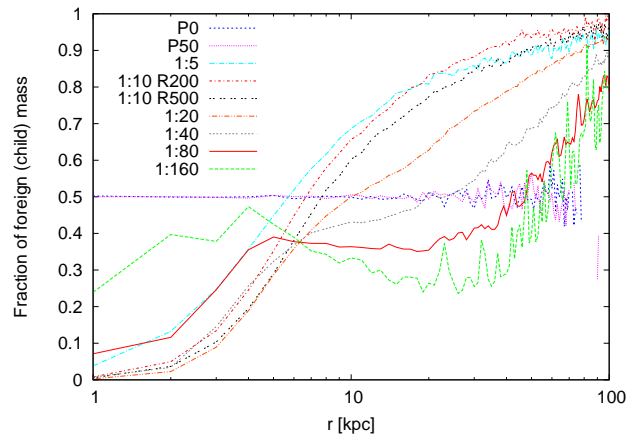}
 \caption{Origin of the mass in the merger remnant at
   $T=10$\,Gyr. Along the y-axis we present the fraction of mass that
   originates from the child galaxies at distance $r$ from the center
   of the merger remnant. We binned the mass in 1\,kpc bins.  When the
   fraction drops to 0 then there is no contribution in mass from the
   child galaxies.}
 \label{fig:PlotRadDistr}        
\end{figure}

The effects described above influence where the mass of the child
galaxies is accreted in the merger remnant.  In
Fig.\,\ref{fig:PlotRadDistr} we present the fraction of baryonic mass
that originates from the child galaxies as a function of the distance
to the center of the merger remnant.  The curves are constructed by
binning the particles in 1\,kpc bins and compute the fraction of mass
that was brought in by the child galaxies. The curves representing the
major mergers (solid-red for model P0 and dashed-green for P50) are
horizontal and at 0.5 in Fig.\,\ref{fig:PlotRadDistr}, which indicates
that they mix homogeneously. Minor mergers experience a different
mixing; the precise effect depends sensitively on the number of
infalling galaxies.

We take a closer look at the 1:5 to 1:20 mass ratios mergers. These
configurations behave as expected \cite{2012arXiv1206.5004H}; the
fraction of child material is small in the core of the merger remnants
and high in the outskirts ($r\apgt 10$\,kpc).  While the mass ratio
continues to decrease the fraction of foreign material in the
outskirts also decreases; until the majority of outskirt
material is original for the most extreme mass ratio (1:160).  We have
to note here that for the most extreme mass ratio's the merger process
has not completed at $T=10$\,Gyr, which is noticeable in
Fig.\,\ref{fig:PlotRadDistr} by the mean line appearing smaller than
0.5.  In the central portion ($r\aplt 7$\,kpc) of the most extreme
mass ratio mergers the opposite happens, in the sense that the
contribution of foreign material is actually increasing (in particular
noticeable for the 1:160 mass ratios).

\begin{figure*}
 \includegraphics[width=1.4\columnwidth, angle=-90]{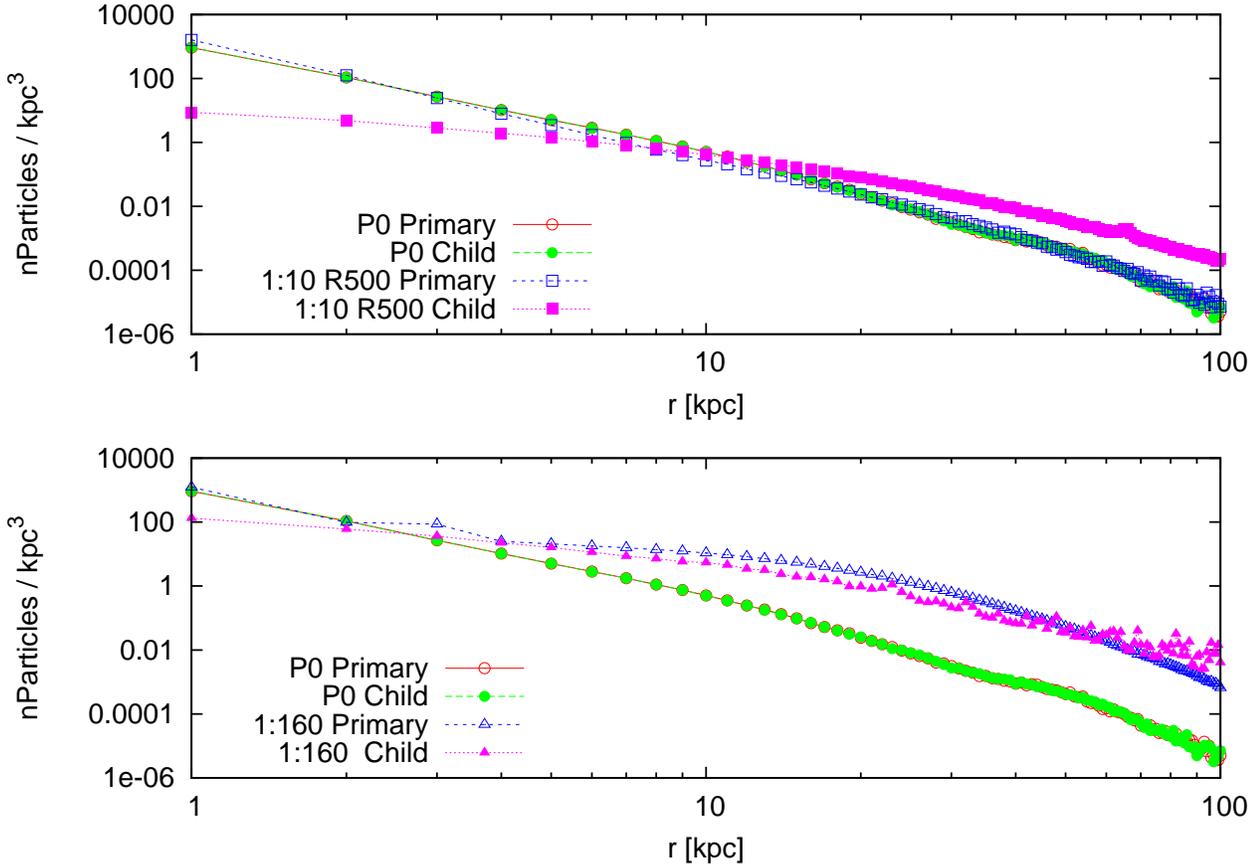}
 \caption{Particle distribution as function of distance from center,
   normalized to the volume size. The x-axis shows the distance to the
   remnants core, the y-axis the normalized particle count. This plot
   is similar as a density distribution, but now split into two lines
   indicating the origin of the particles (mass).  The top and bottom
   panels both show the same averaged major merger result (P0).  In
   addition the top panel shows the average result for the 1:10
   $R=500$ merger and the bottom panel the averaged result for the
   1:160 mergers. Visible is that the 1:10 and major merger show
   different behaviour, where the major merger shows perfect mixing,
   the 1:10 shows a clear distinction between the primary and child
   galaxy mass distribution. The 1:160 however, shows a distribution
   that is closer to the major merger distribution than to the 1:10
   distribution, indicating that 1:160 shows more signs of particle
   mixing.}
 \label{fig:histRadDistrDens}        
\end{figure*}

In Fig.~\ref{fig:histRadDistrDens} we present the number density of
the foreign and native particles as a function of the radius.  We do
this for a mass ratio of 1:1, 1:10 and 1:160.  As we already discussed
in relation to Fig.~\ref{fig:PlotRadDistr} the mass in the major
merger remnant is evenly distributed between the primary and the
infalling galaxy.  In the 1:10 merger (top panel) the two curves for
the foreign material and for the native material are clearly separated,
in contrast to the lines for the major merger. This indicates that the
native material is distributed differently than the foreign
material. The same effect is noticeable in the 1:160 merger (lower
panel in Fig.~\ref{fig:histRadDistrDens}), but less pronounced.  This
indicates that in the 1:160 merger the material of the primary and
child galaxies is mixed more evenly and approaches a mass distribution
similar to that of the major mergers, whereas in the 1:10 merger the
native and foreign material are distributed quite differently.

The results presented in the previous \S\ indicate that for mass
ratios 1:1 to 1:20 the accretion behaves according as if they grew
from inside-out; the material of the child galaxies is stripped and
deposited in the outside of the primary galaxy
\citep{2012arXiv1206.5004H}.  When the number of child galaxies
increases beyond 20 this behavior changes in that the mixing becomes
more homogeneous.  This is caused by the self-interactions of the
child galaxies with the nucleus of the primary. These
self-interactions prevent the merger galaxy from settling to
equilibrium and cause the core to heat up.

We theorize that the continuous bombardment of minor galaxies causes the merger remnant
to remain quite dynamic which reduces the dynamical friction. This in
itself allows new incomers to penetrate deeper into the parent galaxy.
This effect mimics the process of violent relaxation
\citep{1967MNRAS.136..101L}. This effect becomes more efficient when
the number of child galaxies increases.  During the in-fall they
interact with other infalling children before they break-up, after
which their mass is distributed evenly in the merger remnant.  The
interactions are not strong enough to permanently eject a child.  When
the child galaxies are more massive 1:5 to 1:20 mergers self
interactions between the children is negligible and they tend to be
disrupted upon the first pericenter passage of the merger product.  As
a result their material tends to be deposited in the outskirts.

The net growth in size observed in our simulations exceeds those
reported in \citep{2009ApJ...706L..86N,2012MNRAS.422.1714N}.  The
initial conditions of \citep{2009ApJ...706L..86N,2012MNRAS.422.1714N},
however, deviates from ours in the sense that they adopt a more
realistic mass and shape distribution than in our more theoretical
approach. As a consequence the results cannot be compared trivially.
The more efficient growth in our simulations is a result in the
distribution of the minor galaxies, and not by the choice of their
densities; the effect of the child densities is negligible (see
Appendix\,\ref{Sect:App:Density}).  Our results are consistent with
those of \cite{2012arXiv1206.1597H, 2012MNRAS.tmp...42O} for mass
ratios $<1:20$, in the sense that the growth in size is sufficiently
efficient to explain the observational results when the number of
infalling galaxies $>$ 5 (and consequently a mass ratio $<1:5$).  For
more extreme mass ratios $\geq 1:20$ we find a gradual change in the
behavior. This regime was not simulated by
\cite{2012arXiv1206.1597H} and \cite{2012MNRAS.tmp...42O}.

\section{Conclusion}

We have studied dry minor mergers of galaxies in order to understand
the observed growth in size without much increasing the mass of
compact massive galaxies. The simulations ware performed the Bonsai
GPU enabled tree code for $N$ op to 17.6\, million particles.  Our
simulations start with a major galaxy and a number of minor galaxies
in a kinematically cold (and warm) environment. The total mass of the
child galaxies equals that of the primary. The simulation were
performed for 10\,Gyr, and we study the size growth of the merger
remnant. We demonstrate that mergers with a mass ratio 1:5 to 1:20
satisfactory explain the observed growth in size of compact galaxies.
The growth of the merger remnant is always at least a factor of two
higher than in the case of a major merger. This result is robust
against variations in the initial density of the child galaxies.  The
mass of the minor galaxies tend to be accreted to the outside of the
merger remnant.  As a consequence the core of the merger remnant,
formed by the original primary galaxy, will hardly be affected by the
merging process.  This 'inside-out-growth' is consistent with previous
studies \citep{2012arXiv1206.5004H}.

If the number of child galaxies exceeds 20 the behavior of accretion
changes, in the sense that the minor galaxies tend to accumulate in
the central portion of the merger remnant.  Evidence for this
'outside-in' growth is present in the density and velocity profiles of
the merger remnant. The outside-in growth is mediated by self
interactions among the child galaxies before they dissolve in the
merger remnant.

We conclude that the observed large massive elliptical galaxies can be
evolved from compact galaxies at $z\apgt 2$ if they have grown in mass
by accreting 5-10 minor galaxies with a mass ratio of 1:5 to 1:10.
The majority of accreted mass will be deposited in the outskirts of
the merger remnant.

\section*{Acknowledgments}

We thank Marijn Franx and Steven Rieder for discussions.  This work is
supported by NWO grants (grants \#643.000.802 (JB), \#639.073.803
(VICI), and \#612.071.305 (LGM)) and by the Netherlands Research
School for Astronomy (NOVA).

\bibliographystyle{mn2e.bst}

\appendix

\section{Resolution effects}

\begin{figure}
 \includegraphics[width=0.7\columnwidth, angle=-90]{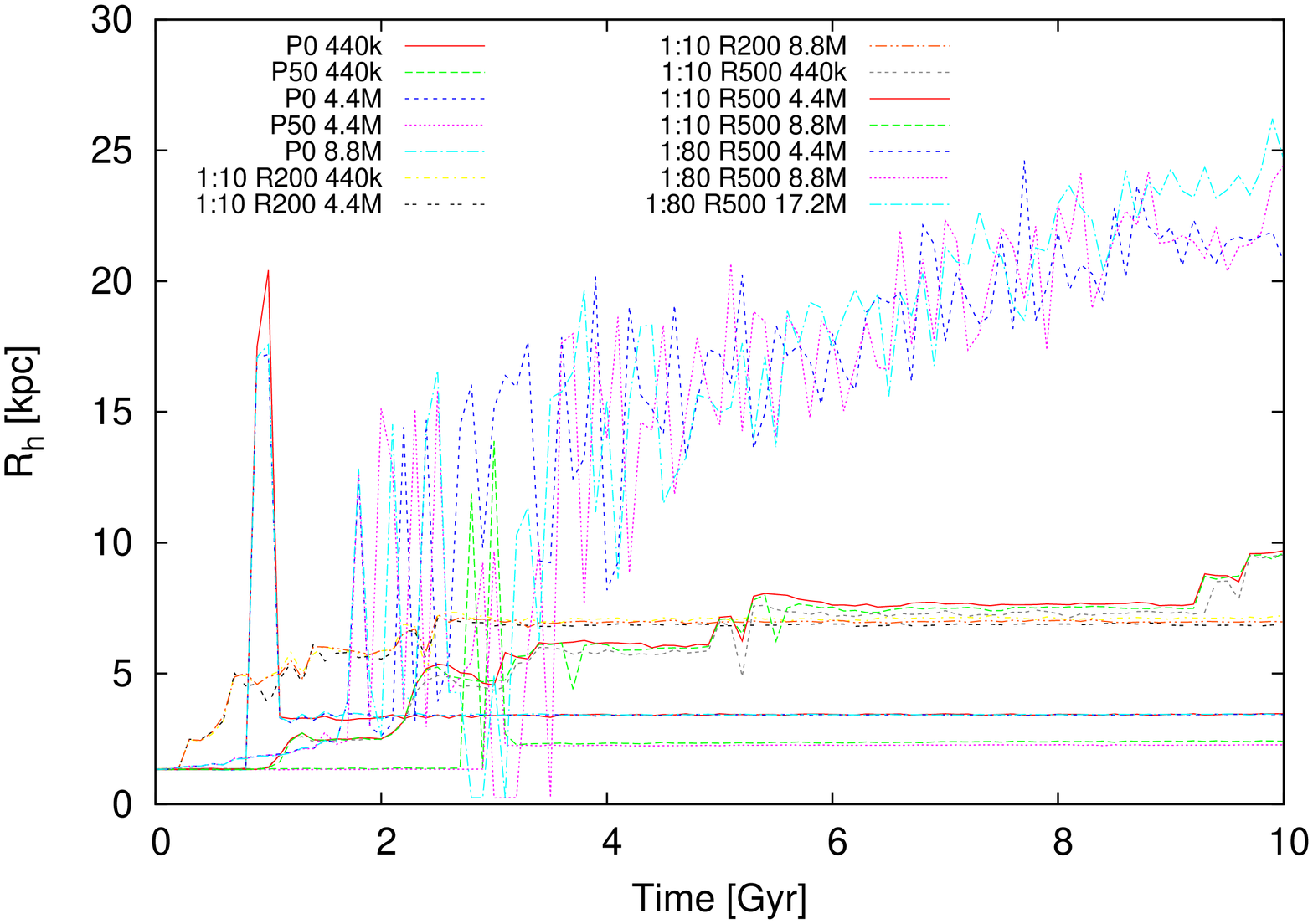}
 \caption{$R_{h}$ versus time for a subset of configurations. Each configuration is run
  with three different resolutions to test dependence on the number of particles used. 
  We present the results for two major merger configurations ($P=$ 0 and 50\,kpc), 
  for two 1:10 minor  merger configurations ($R=200$ and ($R=500$) and one 1:80 merger configuration.}
 \label{fig:resolutionOverview}        
\end{figure}

As discussed in Sect.~\ref{Sect:ConstrainingIC} we perform a range of simulations with
various resolutions for different major and minor merger setups. Depending on the number of child galaxies we 
use low (LR) and high (HR) resolution configurations (see Tab.~\ref{Tab:GalProgen}).
In the LR configurations
we adapted $N=4.4 \times 10^5$ particles of which half belongs to the
primary and the other half to the child galaxies. For the HR simulations we have
$N=4.4 \times 10^6$. To verify that the chosen resolutions 
are sufficient we ran a subset of our configurations with even higher resolutions.
The results can be seen in Fig.~\ref{fig:resolutionOverview}.
Here we present the size evolution of two major merger configurations, a 1:10 merger
with $R=200$\,kpc and $R=500$\,kpc and one of the 1:80 mergers.
Each configuration is run using the default resolution,
as specified in Tab.~\ref{Tab:ICs}, and using two higher resolutions
For the 1:80 simulation this translates to $N=4.4 \times 10^6$, 
$N=8.8 \times 10^6$ and $N=17.2 \times 10^6$. For the other 
configurations we use $N=0.44 \times 10^6$, $N=4.4 \times 10^6$ 
and $N=8.8 \times 10^6$. For each of the resolutions we 
checked if the chosen time-step and softening was sufficient 
to keep the model stable in isolation
using the method described in Sect.~\ref{Sect:ConstrainingIC}.
In all but the 1:80 configurations there is little to no difference
in the size of the merger remnant over the course of the 
simulation. Indicating that our choice of $N$ is 
sufficiently large. 
The large amount of child galaxies in the 1:80 
configurations cause child galaxies to interact
with each other and not directly merge with the 
primary galaxy. The galaxies keep flying 
in and out of the remnant and makes it harder 
to determine $R_{h}$ of the merger remnant which
causes the large fluctuation in size of the 1:80 merger
models. However, the trend of the $R_{h}$ increase is 
independent of the used resolution. 
As final test we performed the 1:80 configuration with
$N=4.4 \times 10^5$ particles (not plotted) which 
quantitatively gives the same result, indicating that
$N=4.4 \times 10^6$ is adequate for the 1:80 configurations.

\section{The effect of child density}
\label{Sect:App:Density}

As described in Sect.~\ref{Sect:ConstrainingIC} we do not 
change the density of the child galaxies when generating the initial conditions. 
Instead we change the cut-off radius to be a factor, $f$, smaller than that of the 
primary galaxy. This $f$ scales with the mass ratio of the child galaxy.
For example the cut-off radius of a child galaxy with mass ratio 
1:10 is 10 times smaller than that of the primary. Together 
with the cut-off radius the mass is also reduced by a factor 10
giving the child galaxies a density lower than that of the 
primary galaxy (\S\,\ref{Sect:ConstrainingIC}). 

To test the effect of the density on the size growth of
the merger product we generated two new 1:10 child galaxies.
In addition to the {\tt standard} child galaxy we created a {\tt compact} 
and a {\tt supercompact} child galaxy. These galaxies are generated
such that their densities are higher than that of the {\tt standard} 
child galaxy. The galaxy properties are presented in Tab.~\ref{Tab:DensChildIC}.

\begin{table}
\begin{tabular}{lllll}
\hline
Model & Bulge mass & Cut-off & $R_h$ & $\rho$ \\
       & [$\Msun$] & [kpc] & [kpc] & [$10^{10}M_{\sun}/kpc^3]$ \\
\hline
  Primary             & $2\times10^{11}$  & 100 & 1.36 & 1232\\
  {\tt standard}      & $2\times10^{10}$  & 10  & 1.26 & 142\\
  {\tt compact}       & $2\times10^{10}$  & 5.1 & 0.68 & 307\\
  {\tt supercompact}  & $2\times10^{10}$  & 0.9 & 0.25 & 4609\\
\hline

\end{tabular}
\caption{
Properties of the extra 1:10 child configurations. The first column
indicates the model, either the primary or one of the three child configurations.
The second column indicates the bulge mass. The third the cut-off radius for the 
dark matter halo, which is used to configure the galaxies.
The fourth (fifth) column indicates the half-mass radius (density)
of the galaxy in isolation at T=1\,Gyr.
}
\label{Tab:DensChildIC}
\end{table}

With the new child galaxies in place we selected one of 
the 1:10 $R=500$\,kpc merger configurations and ran this
realization with the new child galaxies. Each model is evolved
for 10\,Gyr using high resolution ($N=4.4 \times 10^6$).

\begin{figure}
 \includegraphics[width=0.75\columnwidth, angle=-90]{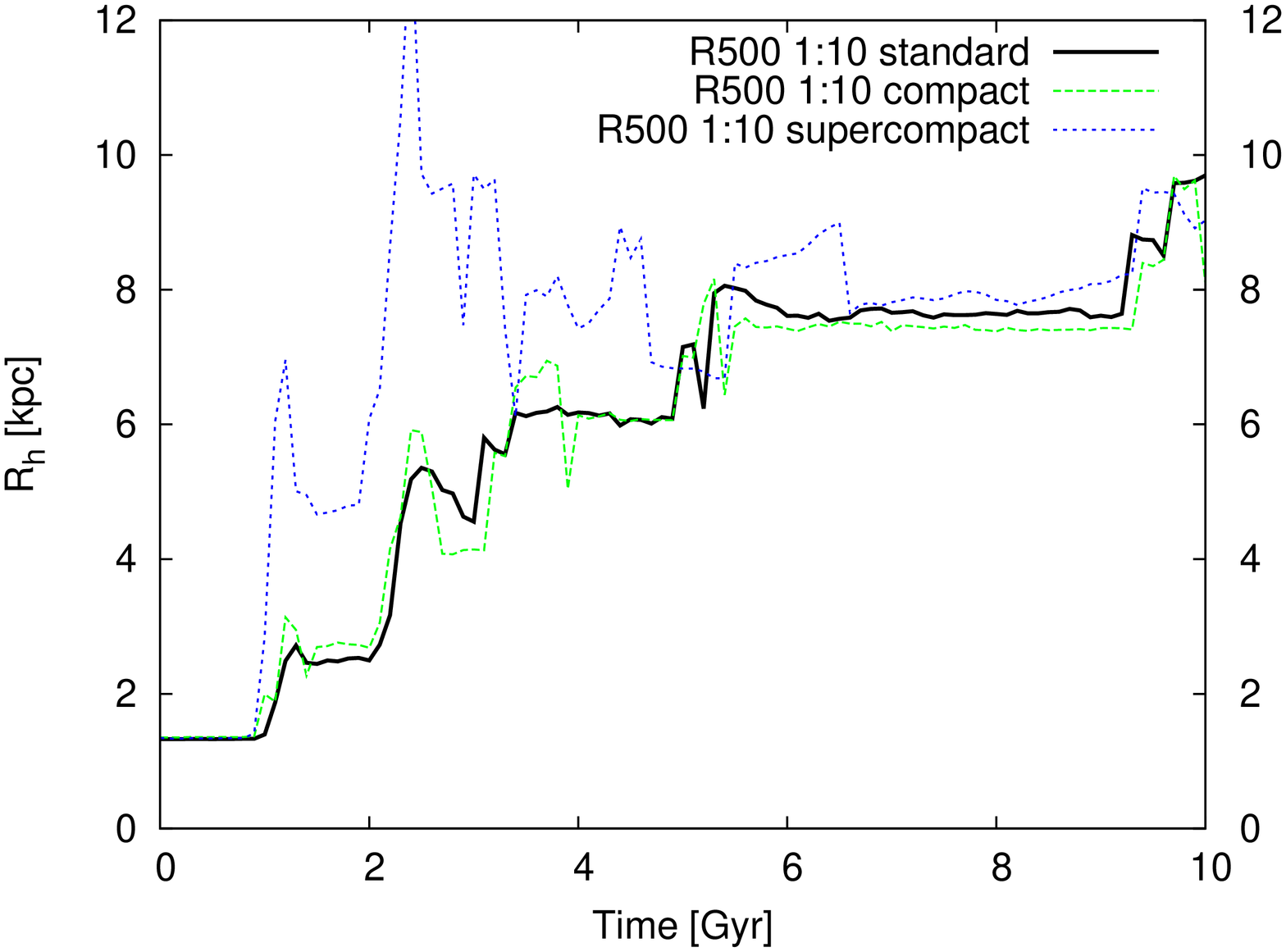}
 \caption{Size evolution. The configurations are based on the same primary galaxy
          and Plummer distribution, but the child galaxies have different size and
          density properties.}
 \label{fig:densityCheckRadius}        
\end{figure}

The effect of the child density on the size growth of the merger
remnant is presented in Fig.~\ref{fig:densityCheckRadius}, where we show 
$R_{h}$ as a function of time. The {\tt standard} and {\tt compact} 
configurations show similar evolution. The {\tt supercompact}
configuration shows a different behaviour during the first 5\,Gyr of the 
simulation. This is an artifact of the way we determine the 
size of the merger remnant which does not work perfect in this 
particular configuration (see \S\,\ref{Sect:Results}). The compactness of the galaxies causes 
the cumulative mass-profile to be irregular compared to the simulations
with less dense child galaxies. As a consequence we are not always 
able to correctly measure the position in the profile that indicates the end of the 
merger remnant. After 5\,Gyr the procedure works correct, because at 
this point the child galaxies have been involved in so many interactions
that the mass-profile is smooth again. When the method starts working
again we notice that all three simulations show similar size growth, independent 
of their density.

\subsection{Circular velocity}

\begin{figure}
 \includegraphics[width=0.75\columnwidth, angle=-90]{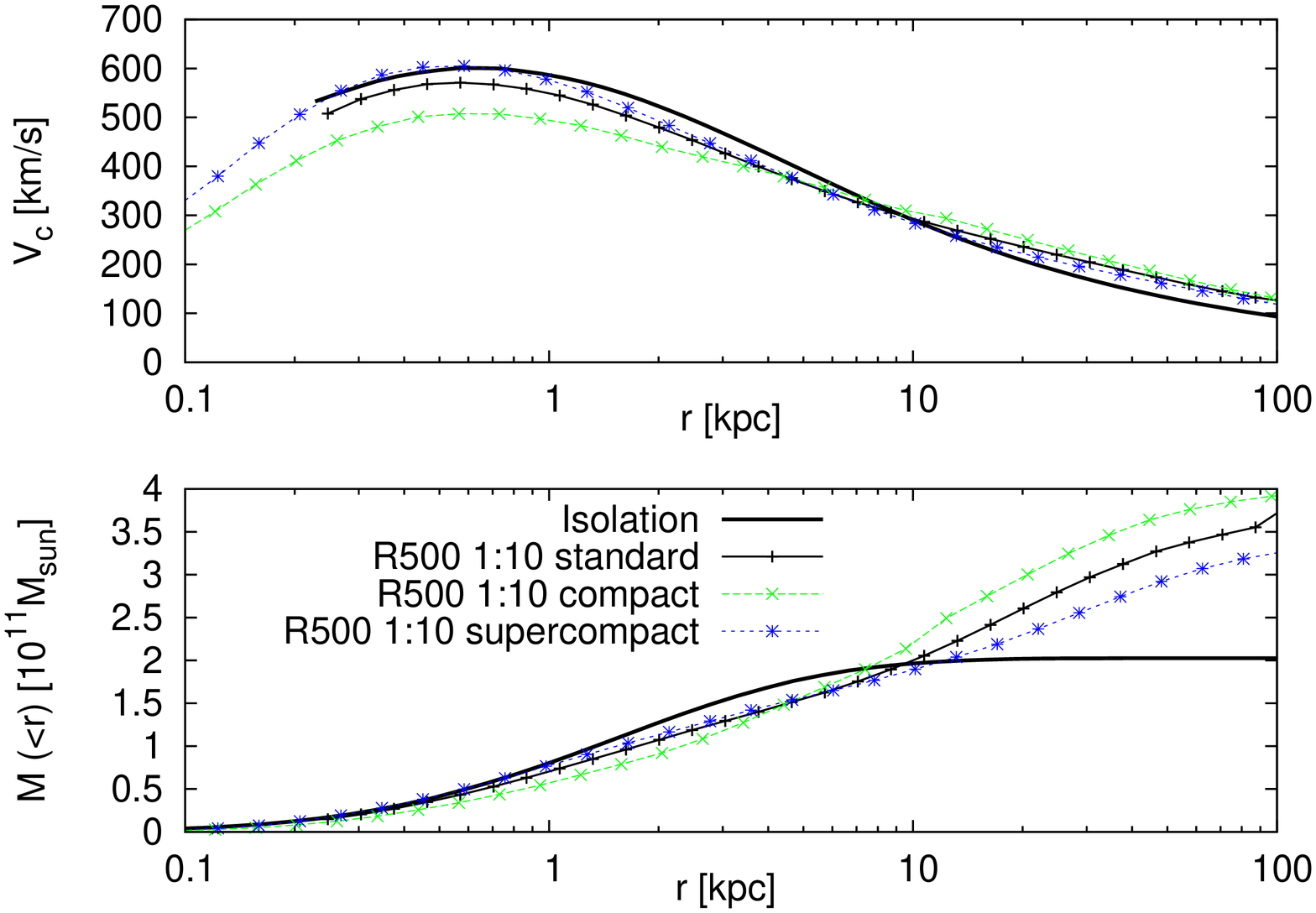}
 \caption{Circular velocity (top) and cumulative mass (bottom) after 10\,Gyr for
   different merger configurations. The mergers are based on the same Plummer distribution
   but the child galaxies have different densities.
   The velocity and mass profile of the isolated model
   are indicated with the thick solid line. }
 \label{fig:densityCheckCircVelocity}        
\end{figure}

The circular velocity of the standard and high density
child simulations is presented in Fig.~\ref{fig:densityCheckCircVelocity} (top) 
we notice that the density of the child galaxy does affect the results. The simulation with the most dense child 
galaxies (striped line with stars) has a circular velocity similar to that of the 
isolated galaxy (thick solid line). The {\tt compact} configuration has a circular velocity
that is $\sim$ 100 km/s lower than the {\tt supercompact} configuration.
The {\tt standard} configuration has a circular velocity in between the 
results of the  {\tt compact} and {\tt supercompact} models. There is no
clear trend in the effect the child density has on the circular velocity.
A similar effect can be seen in the cumulative mass distribution 
of Fig.~\ref{fig:densityCheckCircVelocity} (bottom). 
The difference between the runs is caused by the changed merger history.
Even though the configurations are based on identical Plummer realisations 
the difference in density causes the merger history to be altered. 
Tidal stripping has less effect on the more compact galaxies when they 
move through the dark-matter halo of the primary than it has on the {\tt standard}
galaxy.  Instead the higher density causes these galaxies to stay tightly bound 
resulting in interactions between other children and the core of the primary galaxy. 
Because of these interactions some of the child galaxies will be flung from 
the galaxy during the first passage 
instead of merging with the primary.

\begin{figure}
 \includegraphics[width=0.75\columnwidth, angle=-90]{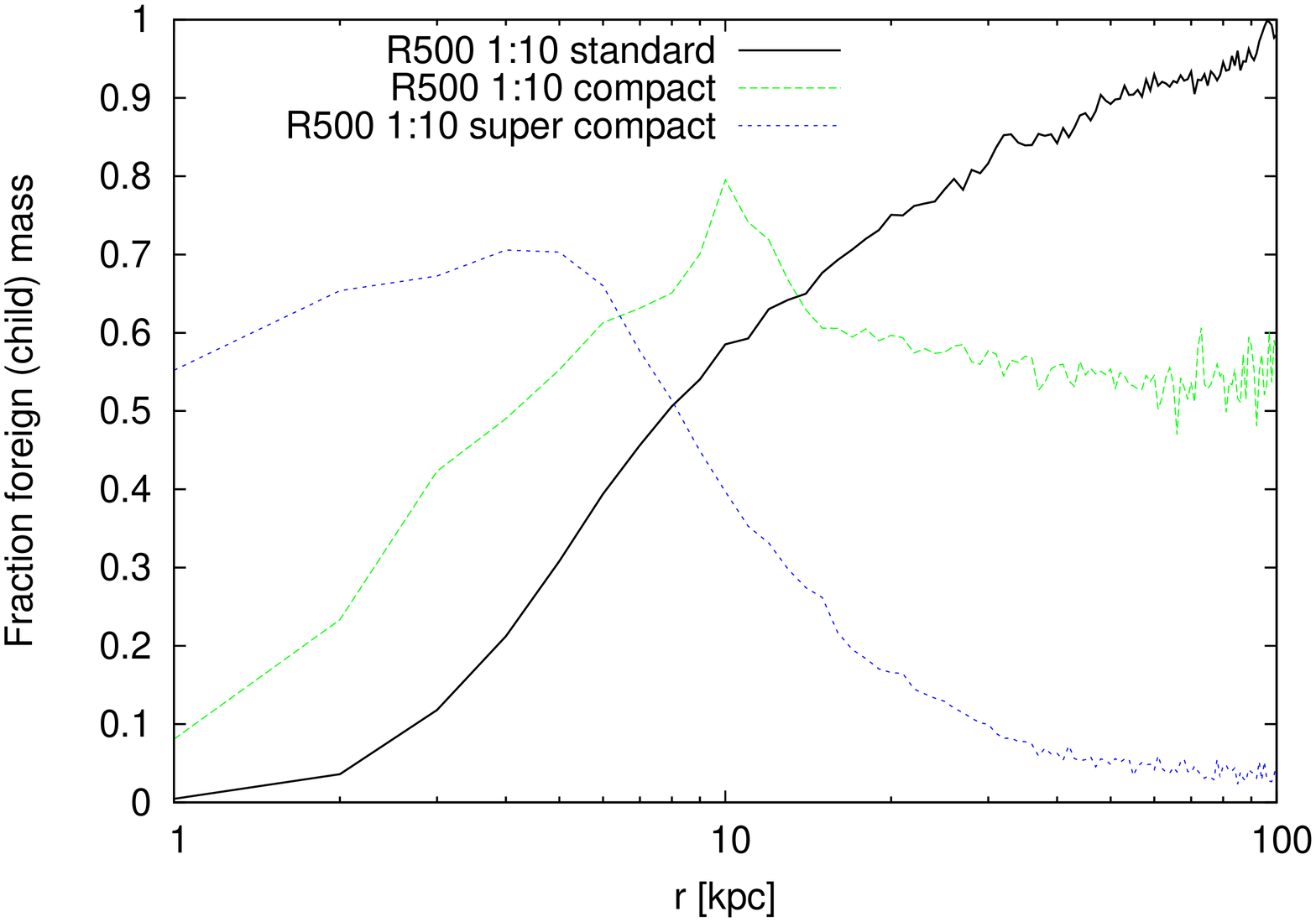}
 \caption{Origin of the mass in the merger remnant. Shown on the
   y-axis the fraction of mass that originates from the primary
   galaxy at a certain radius from the center. The mass is
   binned in 1\,kpc sized bins. If the fraction is 0 then
   there is no contribution in mass from the child galaxies.}
 \label{fig:densityCheckRadDistr}        
\end{figure}

In Fig.~\ref{fig:densityCheckRadDistr} we present the distribution of 
accreted material throughout the merger remnant. In this figure we notice
large differences between the different configurations caused by the 
differences in child density. The solid line represents the {\tt standard}
simulation where the mass is accreted
on the outside of the primary (outside-in formation). For the denser children
the mass is accreted closer to the core of the primary galaxy.
This is the result of the less effective tidal-stripping described
in the previous \S. The dense child galaxies are able to progress further 
to the core of the merger remnant before they break up and 
are accreted. While the {\tt standard} galaxies loose much of their 
material outside the core.
This effect is best indicated by the {\tt supercompact} configuration. 
In the merger remnant of this simulation most of the mass in the inner 
8\,kpc originates from the child galaxies. While in the other configurations
most of the mass in this inner part of the merger remnant originates from the primary.

Concluding we can say that the density has no effect on the size
of the merger remnant, but it does have an effect on how the remnant
is build-up. With either mass being accreted on the outskirts (low density children)
or in the inner parts of the merger remnant (high density children).

\label{lastpage}

\end{document}